\pdfoutput=1

\documentclass[11pt,twoside,a4paper,cmspaper,final,collab]{cms-tdr}
\def\svnVersion{467937P}\def\cmsCernNoTag{CERN-EP-2017-249}\def\cmsCernDate{\today}\def\cmsMessage{Published in the Journal of High Energy Physics as \href{http://dx.doi.org/10.1007/JHEP07(2018)032}{\doi{10.1007/JHEP07(2018)032.}}}

\begin{document}\cmsNoteHeader{FSQ-16-008}

\hyphenation{had-ron-i-za-tion}
\hyphenation{cal-or-i-me-ter}
\hyphenation{de-vices}
\RCS$HeadURL: svn+ssh://svn.cern.ch/reps/tdr2/papers/FSQ-16-008/trunk/FSQ-16-008.tex $
\RCS$Id: FSQ-16-008.tex 463842 2018-06-08 12:12:02Z rgupta $
\newlength\cmsFigWidth
\ifthenelse{\boolean{cms@external}}{\setlength\cmsFigWidth{0.85\columnwidth}}{\setlength\cmsFigWidth{0.4\textwidth}}
\ifthenelse{\boolean{cms@external}}{\providecommand{\cmsLeft}{top\xspace}}{\providecommand{\cmsLeft}{left\xspace}}
\ifthenelse{\boolean{cms@external}}{\providecommand{\cmsRight}{bottom\xspace}}{\providecommand{\cmsRight}{right\xspace}}
\newcommand{\pTmumu}{\ensuremath{\pt^{\mu\mu}}\xspace}
\newcommand{\x}{\ensuremath{\phantom{0}}}
\newcommand{\y}{\ensuremath{\phantom{.}}}

\cmsNoteHeader{FSQ-16-008}

\title{Measurement of the underlying event activity in inclusive Z boson production in proton-proton collisions at $\sqrt{s} = 13\TeV$}

\date{\today}

\abstract{
This paper presents a measurement of the underlying event activity in proton-proton collisions at a center-of-mass
energy of 13\TeV, performed using inclusive Z boson production events collected with the CMS experiment
at the LHC. The analyzed data correspond to an integrated luminosity of 2.1\fbinv.
The underlying event activity is quantified in terms of the charged particle multiplicity, as well as of the scalar sum of the charged particles' transverse momenta in different topological regions defined with respect to the Z boson direction.
The distributions are unfolded to the stable particle level and compared with predictions from various Monte Carlo event generators, as well as with similar CDF and CMS measurements at center-of-mass energies of 1.96 and 7\TeV respectively.
}

\hypersetup{%
pdfauthor={CMS Collaboration},%
pdftitle={Measurement of the underlying event using inclusive Z boson process in proton-proton collisions at sqrt(s) = 13 TeV},%
pdfsubject={CMS},%
pdfkeywords={CMS, physics, underlying event}}

\maketitle

\section{Introduction}

The production of particles in a hadron-hadron collision includes contributions from parton-parton scatterings, initial-state
radiation (ISR), final-state radiation (FSR), and beam-beam remnant (BBR) interactions. The large parton densities accessible
in proton-proton (pp) collisions at the CERN LHC result in a significant probability of more than one
parton-parton scattering in the same pp collision, a phenomenon known as multiple parton interactions (MPI).
The combination of particle production from MPI (excluding the parton-parton scattering with the highest momentum transfer) and BBR interactions is commonly called the underlying event (UE). The UE usually produces particles at low transverse momentum ($\pt$) that cannot be experimentally distinguished from the particles produced from ISR and FSR. These processes cannot be completely described by perturbative
quantum chromodynamics (QCD) calculations, and require phenomenological models, whose parameters are tuned by means of fits to data.

The experimental measurement of the UE is often based on a process that defines the scale of the
hardest parton-parton scattering, along with a phase space region with enhanced sensitivity to particle production associated with the UE activity. A number of measurements~\cite{uecms1,uecms2,uedycms,uealice,ueatlas1,ueatlas2,ueatlas3,ueatlas4,uecdf}
have been performed by the Tevatron and LHC experiments at various center-of-mass energies, ranging from
0.3\TeV to 13\TeV, and using a variety of hard processes including events with high-$\pt$ charged particles or jets, Z+jets, and $\ttbar$+jets.
Measurements of the UE associated with different hard processes are useful to test the level of universality of the underlying MPI dynamics. Events with a harder scale are expected to correspond, on average, to proton-proton interactions with a smaller impact parameter and therefore with more MPI~\cite{Sjostrand:1986ep}. Such increased UE activity is observed to plateau at high energy scales, which indicates that the smallest impact parameters have been reached and hence maximum matter overlap in the pp collision~\cite{Frankfurt:2011}.

This paper presents a measurement of the UE activity based on events with inclusive $ \PZ\rightarrow \mu^{+}\mu^{-}$ production at $\sqrt{s} = 13\TeV$. Underlying event measurements based on Z boson production have been carried out previously at $\sqrt{s} = 1.96\TeV$~\cite{uecdf} and 7\TeV~\cite{uedycms,ueatlas4} by Tevatron and LHC experiments. Z boson production is a process with a clean experimental signature and well understood theoretically, allowing clear identification of the UE activity. Measurements with Z bosons also make it possible to partially distinguish the MPI and ISR/FSR contributions~\cite{uedycms,Bansal:2016iri}. In this paper, the properties of the UE are measured as a function of conventional observables related to the impact parameter of the pp collision, such as the number of charged particles and the
scalar sum of their $\pt$. The data are corrected for detector effects and compared to Monte Carlo (MC)
event generators, as well as with earlier results at $\sqrt{s} = 1.96\TeV$~\cite{uecdf} and 7\TeV~\cite{uedycms}.

The outline of the paper is as follows. Section~\ref{sec:mc} describes the data and simulated samples used for the
validation and unfolding studies. Section~\ref{sec:cmsdetector} gives a brief description
of the CMS detector, whereas Section~\ref{sec:EnTsel} describes the event and track selection criteria, and the observables used for quantification of the UE. The unfolding procedure and systematic effects are discussed in Section~\ref{sec:Unfold}, and the final results are
presented in Section~\ref{sec:results}. Finally, the analysis is summarized in Section~\ref{sec:summary}.

\section{Data and simulated samples}
\label{sec:mc}

The analysis is performed on a sample of pp collisions at $\sqrt{s} = 13\TeV$, corresponding to an
integrated luminosity of 2.1\fbinv. Data were collected with the CMS detector in 2015 when the average
number of inelastic collisions per bunch crossing (pileup) was about 20.

For the evaluation of the event and track selection efficiencies, signal and background processes are simulated at next-to-leading order (NLO) accuracy with \MCATNLO 2.2.2~\cite{Alwall:2014hca} and, for single top production, with \POWHEG 2.0~\cite{Frixione:2007,MINLO}. To study the model dependence, the Z+jets events are also simulated at leading order (LO) with \MADGRAPH{5} 2.2.2~\cite{Maltoni:2003,Alwall:2011} combined with \PYTHIA{8}~\cite{Sjostrand:2007gs} using the CUET8PM1~\cite{Khachatryan:2015pea} tune. Diboson (WW, WZ and ZZ) as well as multiple-jet production, via strong interaction processes, are
generated at LO with \PYTHIA{8} standalone. The NNPDF3.0~\cite{Ball:2014uwa} set is used as the default set of parton distribution functions (PDFs) for all generated LO and NLO samples.

These simulated samples are processed and reconstructed in the same manner as the collision data. The
detector response is simulated in detail by using the \GEANTfour package~\cite{Agostinelli:2002hh}.
The samples include additional pileup pp interactions, with a multiplicity distribution matching that observed in data.

The measured UE distributions are unfolded to correct for detector effects and selection efficiencies, and compared to various MC simulation predictions:

\begin{itemize}
\item
\MADGRAPH~+~\PYTHIA{8}: Z+jets events are generated with \MADGRAPH, followed by parton showering and hadronization with \PYTHIA{8} (CUET8PM1 tune). The \MADGRAPH generator includes up to 4 partons  in the matrix element calculations, while additional jets can be generated by \PYTHIA{8} during parton showering.
\item
\POWHEG~+~\PYTHIA{8}: Z+jets events are produced up to NLO accuracy with the \POWHEG~ `Multiscale-improved NLO' method~\cite{MINLO}. The \PYTHIA{8} generator assumes \pt-ordered parton showers, and the latter are interleaved with MPI. Tune CUET8PM1 is used for hadronization and parton showering. To quantify the effect of MPI, events are also simulated without MPI. To study the impact of color-reconnection (CR) between final state partons, \PYTHIA{8} events are also simulated without CR.
\item
\POWHEG~+~\HERWIG{++}: To further investigate the model dependence, \POWHEG~ \\  events are also hadronized
using \HERWIG{++}~\cite{hpp} with tune EE5C~\cite{Khachatryan:2015pea}. \HERWIG{++}, unlike \PYTHIA{8}, generates angular-ordered parton showers. It simulates MPI according to a model similar to that of \PYTHIA{8}, with tunable parameters for the regularization of the parton-parton cross section at very low momentum transfers, but without the interleaving with parton showers. In most models, the number of MPI follows a Poission distribution with a mean that depends on the overlap of the matter distributions of the hadrons.
\end{itemize}

Monte Carlo events are generated at $\sqrt{s}= 7$ and 13\TeV, as well as for proton-antiproton collisions at $\sqrt{s} = 1.96\TeV$.

\section{The CMS detector}\label{sec:cmsdetector}

The central feature of the CMS apparatus is a superconducting solenoid of 6\unit{m} internal diameter. Within the
solenoid volume are a silicon pixel and strip tracker, a lead tungstate crystal electromagnetic calorimeter,
and a brass and scintillator hadron calorimeter, each composed of a barrel and two endcap sections. Forward
calorimeters extend the pseudorapidity coverage provided by the barrel and endcap detectors. Muons are measured in
gas-ionization detectors embedded in the steel flux-return yoke outside the solenoid, covering the pseudorapidity range $\abs{\eta} < 2.4$, with detection planes based on three technologies: drift tubes, cathode strip chambers, and resistive-plate chambers.

The silicon tracker measures charged particles within the range $\abs{\eta} < 2.5$. It consists of 1440
silicon pixel and 15\,148 silicon strip detector modules and is located in the 3.8\unit{T} field of the superconducting
solenoid. For nonisolated particles of $1 < \pt < 10\GeV$ and $\abs{\eta} < 1.4$, the track resolutions are typically
1.5\% in \pt and 25--90 (45--150)\mum in the transverse (longitudinal) impact parameter \cite{TRK-11-001}. Matching muons to tracks measured in the silicon tracker results in a relative \pt resolution for muons with $20 <\pt < 100\GeV$ of 1.3--2.0\% in the barrel and better than 6\% in the endcaps. The \pt resolution in the barrel is better than 10\% for muons with \pt up to 1\TeV~\cite{Chatrchyan:2012xi}.

A more detailed description of the CMS detector, together with a definition of the coordinate system used and the relevant
kinematic variables, can be found in Ref.~\cite{Chatrchyan:2008zzk}.

\section{Experimental methods}\label{sec:EnTsel}

\subsection{Event selection}

Events are selected online by requiring the presence of at least two isolated muon candidates with $\pt > 17\,(8)$\GeV for the leading (subleading) muon. Offline, events are required to have at least one well-reconstructed vertex~\cite{TRK-11-001} within ${\pm}24\unit{cm}$ of the nominal interaction point along the $z$-direction. At least five tracks are required to be associated with the vertex, which should be at most 2\unit{cm} from the beam axis in the transverse plane. Muons are reconstructed with the particle-flow algorithm~\cite{Sirunyan:2017ulk} and are required to satisfy identification criteria based on the number of hits in the muon detectors and tracker, the transverse impact parameter with respect to the beam axis, and the normalized $\chi^2$ of the global muon track fit. The backgrounds from jets misidentified as muons and from semileptonic decays of heavy quarks are suppressed by applying an isolation condition on the muon candidates. The relative isolation variable, $\mathrm{I}_\text{rel}$, for muons is defined as:
\begin{equation}
\mathrm{I}_\text{rel} = \frac{[\sum \pt^{\smash[b]{\, \text{charged}}} + max(0., \sum E_{\mathrm{T}}^{\smash[b]{\, \text{neutral}}} + \sum E_\mathrm{T}^{\gamma} - 0.5\sum \pt^\mathrm{ PU})]}{\pt^{\mu}}.
\end{equation}

\sloppypar{
Here $\sum E_{\mathrm{T}}^{\smash[b]{\, \text{neutral}}}$ and $\sum E_\mathrm{T}^{\gamma}$ are the sums of the transverse energies of neutral
hadrons and photons, respectively, in a pseudorapidity-azimuth cone of size  $\Delta R \equiv \sqrt{\smash[b]{(\eta^{\mu}-\eta^{\text{neutral},{\gamma}})^{2} + (\phi^{\mu}-\phi^{\text{neutral},{\gamma}})^{2}}} < 0.4 $ around the muon direction. The quantity $\sum \pt^{\smash[b]{\, \text{charged}}}$ represents the $\pt$ sum of the charged hadrons, in the same cone around the muon, associated with the selected vertex. Finally, $\sum\pt^\mathrm{PU}$ is the $\pt$ sum of the charged hadrons, in the same cone around the muon, not associated with the selected vertex. A muon is considered isolated if $\mathrm{I}_\text{rel} < 0.15 $. Misalignment in the detector geometry affects the measurement of muons in a different manner for data and simulation. To account for this effect, different muon momentum corrections~\cite{Bodek:2012id} are applied to data and simulated events.}

{\tolerance=1000
Offline, the leading and subleading muons are required to have a $\pt$ larger than 20 and 10\GeV, respectively, so as to be in the region where the trigger efficiency is highest and $\pt$-independent~\cite{CMS-PAPERS-TRG-12-001}.
These muons are required to be associated to the vertex with the largest
value of the $\pt^{2}$ sum of the tracks belonging to it. Events with two oppositely charged muons are further required to have an invariant mass ($M_{\mu\mu}$) in the window 81--101\GeV. After all the selections, a high-purity sample of Z candidates is extracted with estimated background contributions, mainly from top quark and diboson processes, below 1\%. About 1.3 million Z candidate events are left in the data, which is in agreement within 5\% with the NLO simulation predictions.
\par}

\subsection{Track selection}

All charged particles, except the selected muons, with $\pt > 0.5\GeV$ and $|\eta| < 2 $ are considered for the UE study.
To reduce the number of incorrectly reconstructed tracks, a high-purity reconstruction algorithm~\cite{trk} is used.

The distance of closest approach
between the track and the selected vertex in the transverse plane and in the longitudinal direction are required to
be less than three times the respective uncertainties. These requirements help reduce contamination of secondary tracks
from decays of long-lived particles, photon conversions, and pileup. Tracks with poorly measured momenta are removed
by requiring $\sigma(\pt)/\pt < 5\%$, where $\sigma(\pt)$ is the uncertainty in the $\pt$ measurement. The track selection efficiencies in the data and simulated samples agree within 4--5\%.

These selected charged particle tracks are used to construct the relevant UE observables, namely the particle density and $\Sigma \pt$ density, which are defined as follows:

\begin{itemize}
\item
{Particle density:} The average number of  charged particles in an event per unit $\Delta\eta\Delta\phi$ area.
\item
{$\Sigma \pt$ density:} The average of the scalar $\pt$ sum of all selected charged particles in an event per unit $\Delta\eta\Delta\phi$ area.
\end{itemize}

Here, $\Delta\eta = |\eta^{\Z} - \eta^\text{ch}|$ and $\Delta\phi = |\phi^{\Z} - \phi^\text{ch}|$ are the pseudorapidity and azimuthal separation between each charged particle and the Z boson. In order to enhance the sensitivity to the UE, observables are calculated in different phase-space regions defined with respect to the $\phi$ direction of the Z boson. These regions are classified as:

\begin{itemize}
\item \textit{towards} region: $\Delta\phi< 60^{\circ}$,
\item \textit{transverse} region: $60^{\circ} <\Delta\phi< 120^{\circ}$,
\item \textit{away} region: $\Delta\phi> 120^{\circ}$.
\end{itemize}
The UE observables are studied as a function of the transverse momentum of the dimuon system (\pTmumu).

\section{Unfolding and systematic uncertainties}
\label{sec:Unfold}

In order to compare data and predictions, the UE distributions are corrected to the stable particle level (lifetime $c\tau > 10\unit{mm}$)  with the iterative D'Agostini method~\cite{unfold}, which also accounts for bin-to-bin migrations. In the present analysis, two-dimensional distributions are unfolded with a response matrix constructed from events simulated with \MADGRAPH~+~\PYTHIA{8}.

The unfolded measured distributions may be distorted by a variety of systematic effects, as discussed below.

\begin{itemize}
\item {Model dependence:} The events simulated with \MADGRAPH~+~\PYTHIA{8} reproduce the measured \pTmumu distribution within 10--20\%. The effect of this discrepancy on the final UE distributions is evaluated by reweighting the simulated sample so that it describes the measured \pTmumu distribution. These weights are applied to the response matrix used for the unfolding. The difference between the unfolded distributions with and without these weight factors is 2--5\%. An additional cross-check is performed by using response matrices constructed with events simulated with the \MADGRAPH~+~\PYTHIA{8} and the \MCATNLO~+~\PYTHIA{8} event generators. The difference between the unfolded distributions obtained with the response matrices constructed with these two generators is found to be less than 0.5\%.
\item{Tracking efficiency:} The tracking efficiency is known with an uncertainty of 4\%~\cite{TRK-11-001,CMS:2010mua}. To estimate the effect of this uncertainty on the UE distribution, 4\% of the tracks are randomly removed in the simulated events while constructing the response matrix. The effect on the unfolded distributions is approximately 4--6\%.
\item{Pileup:} Pileup events produce low-\pt particles that can contribute to the UE activity. However, the
effect of pileup is expected to be small in the present analysis because all tracks are required to originate from
the same primary vertex. The effect of pileup is further reduced by the unfolding procedure because the simulated samples
also include pileup. Any possible residual effect is evaluated by varying the pp inelastic cross section used in
the simulation by 5\%. The bias on the unfolded distributions is less than 0.5\%.
\item{Trigger:} The triggers used in the analysis require that the muons be isolated, which may bias the UE
distributions. The effect of this requirement is evaluated by comparing UE distributions obtained with and without
the trigger requirement in the simulation. This affects the results by up to 0.1\%.
\item{Physics background:} The Z boson production events are required to be in the mass window 81--101\GeV. In this region, there is a small (about 0.3\%) contribution of dimuons from diboson and top quark decays. These background processes may bias the UE distributions because of the different event topologies and parton radiation patterns as compared to the Z boson events. The effect of these background processes is evaluated, using simulations, by comparing the UE distributions for the Z-boson events and for the Z-boson events combined with background processes. The UE distributions change by 0.5--1\%.
\item{Muon momentum correction}: The effect of the muon momentum corrections~\cite{Bodek:2012id} is
studied by comparing the corrected data distributions with the ones without corrections. The resulting effect on the
particle density is up to 0.4\%, and up to 0.7\% for the $\Sigma \pt$ density distribution.
\end{itemize}

Table~\ref{tab:sytematics} summarizes the dominant systematic uncertainties in the particle and $\Sigma \pt$ densities. Adding all aforementioned sources in quadrature results in a total systematic uncertainty of 4.8--7.8\%, depending on the UE observable and particular bin.

\begin{table}[htbp]
\centering
\topcaption{\label{tab:sytematics} Summary of the systematic uncertainties in the particle and $\Sigma \pt$ densities.}
\begin{tabular}{l|c} \hline
Observable &  Uncertainty  (\%) \\ \hline \hline
Model dependence & 2--5  \\
Tracking efficiency & 4--6 \\
Pileup & 0.5 \\
Trigger & 0.1 \\
Physics background & 0.5--1\x\y \\
Muon momentum correction & 0.4--0.7 \\
\hline
Total Uncertainty & 4.8--7.8 \\
\hline
\end{tabular}
\end{table}

\section{Results and discussion}
\label{sec:results}

\begin{figure}[htbp]
\begin{center}
\includegraphics[width=0.4\textwidth]{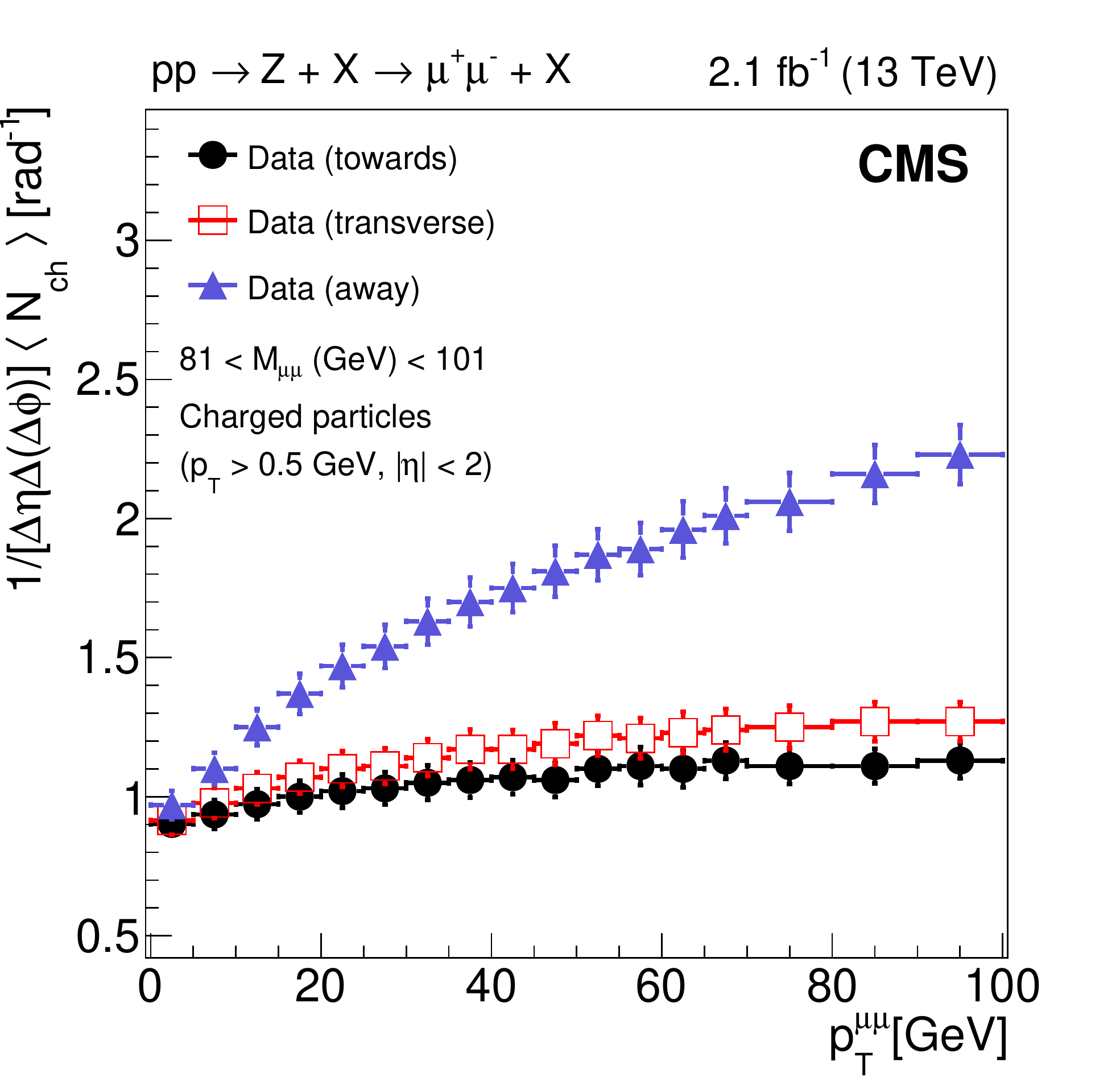} \hfil
\includegraphics[width=0.4\textwidth]{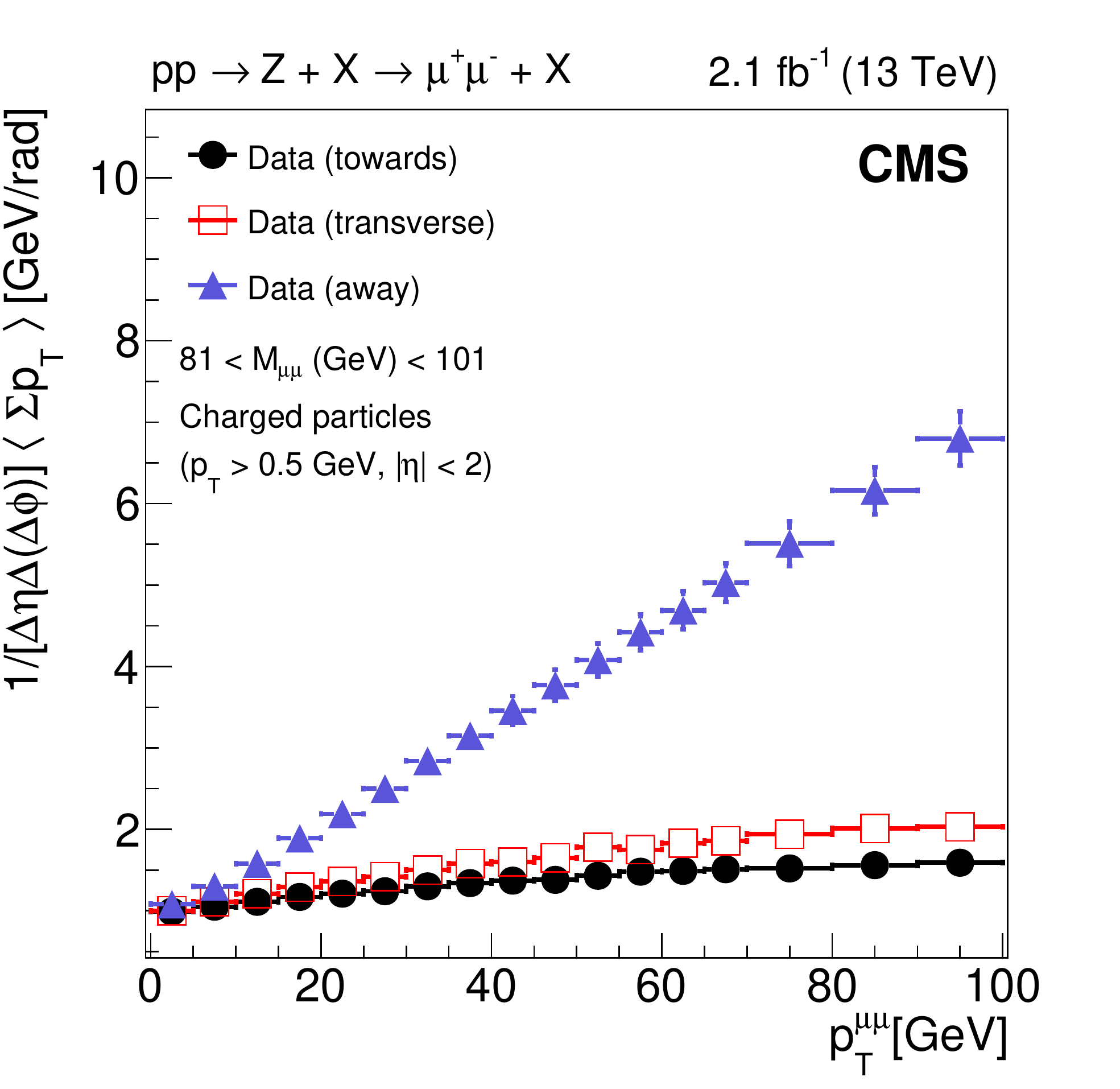}
\end{center}
\caption {{Unfolded distributions of particle density (left) and $\Sigma \pt$ density (right) in Z events, as a function of \pTmumu in the
\textit{towards} ($\Delta\phi< 60^{\circ}$), \textit{transverse} ($60^{\circ} <\Delta\phi< 120^{\circ}$), and \textit{away} ($\Delta\phi> 120^{\circ}$) regions. Error bars represent the statistical and systematic uncertainties added in quadrature.}} \label{fig:compDataMC13allregion}
\end{figure}

Figure~\ref{fig:compDataMC13allregion} shows the comparison of the measured UE activity in the \textit{towards},
\textit{transverse}, and \textit{away} regions. The activity in the \textit{away} region increases sharply with
\pTmumu, but more slowly in the \textit{towards} and \textit{transverse} regions. This is expected as particle production in the \textit{away} region is mostly dominated by the hadronic recoil system, which is highly correlated with \pTmumu. Because of the large spatial separation, the contribution of
the hadronic recoil is small in the \textit{transverse} region, and becomes even smaller in the
\textit{towards} region.
The activity in the three regions becomes similar as \pTmumu approaches zero; this observation again
corroborates the hypothesis that differences in the UE activity for the three regions are due to varying parton radiation
contributions.
Unlike the UE measurement with leading jet/track~\cite{uedycms,ueatlas2}, in the present analysis the UE activity
is not zero when \pTmumu approaches zero.
This behavior reflects the fact that the initial scale in the Z boson events, given by the lepton pair invariant mass in the range 81--101\GeV , is already large enough to determine a significant overlap between the transverse parton densities of the colliding protons, and hence a large number of MPI. From the UE measurements using the leading charged particle (jet) approach~\cite{uedycms,ueatlas2}, it is observed that the MPI contribution reaches its maximal value at an energy scale of 5 (12--15)\GeV. Above this energy, there is a slow rise in the number of particles produced, which is mainly attributed to the increase in the parton radiation contributions. In the present measurement, the minimum scale is set by the dimuon mass (81--101\GeV), which is larger than the energy where the MPI contribution saturates. Therefore, the increase in UE activity with \pTmumu should be mainly ascribed to the rise in the recoil hadronic contribution and associated ISR/FSR~\cite{uedycms}.

\begin{figure}[htbp]
\begin{center}
\includegraphics[width=0.49\textwidth]{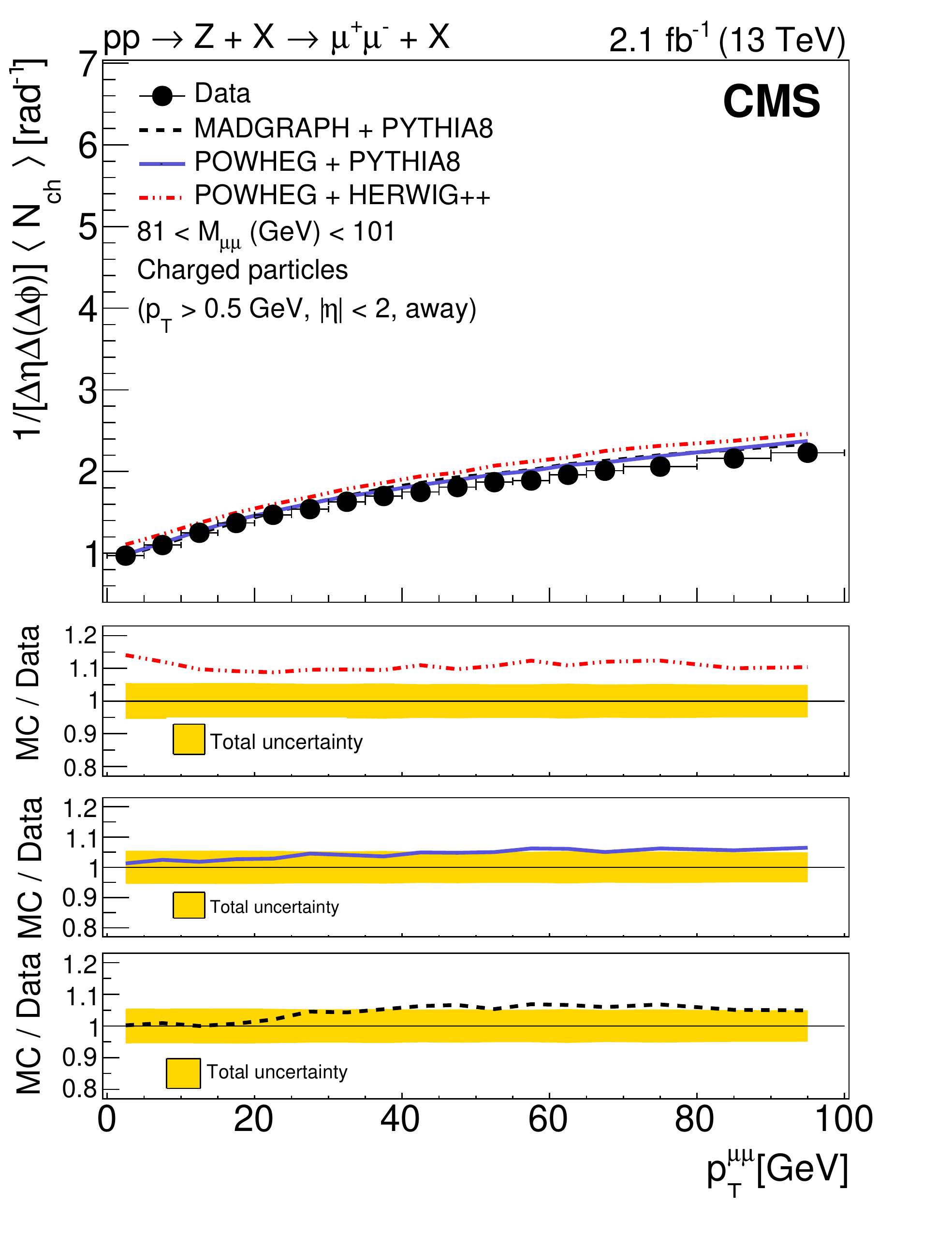} \hfil
\includegraphics[width=0.49\textwidth]{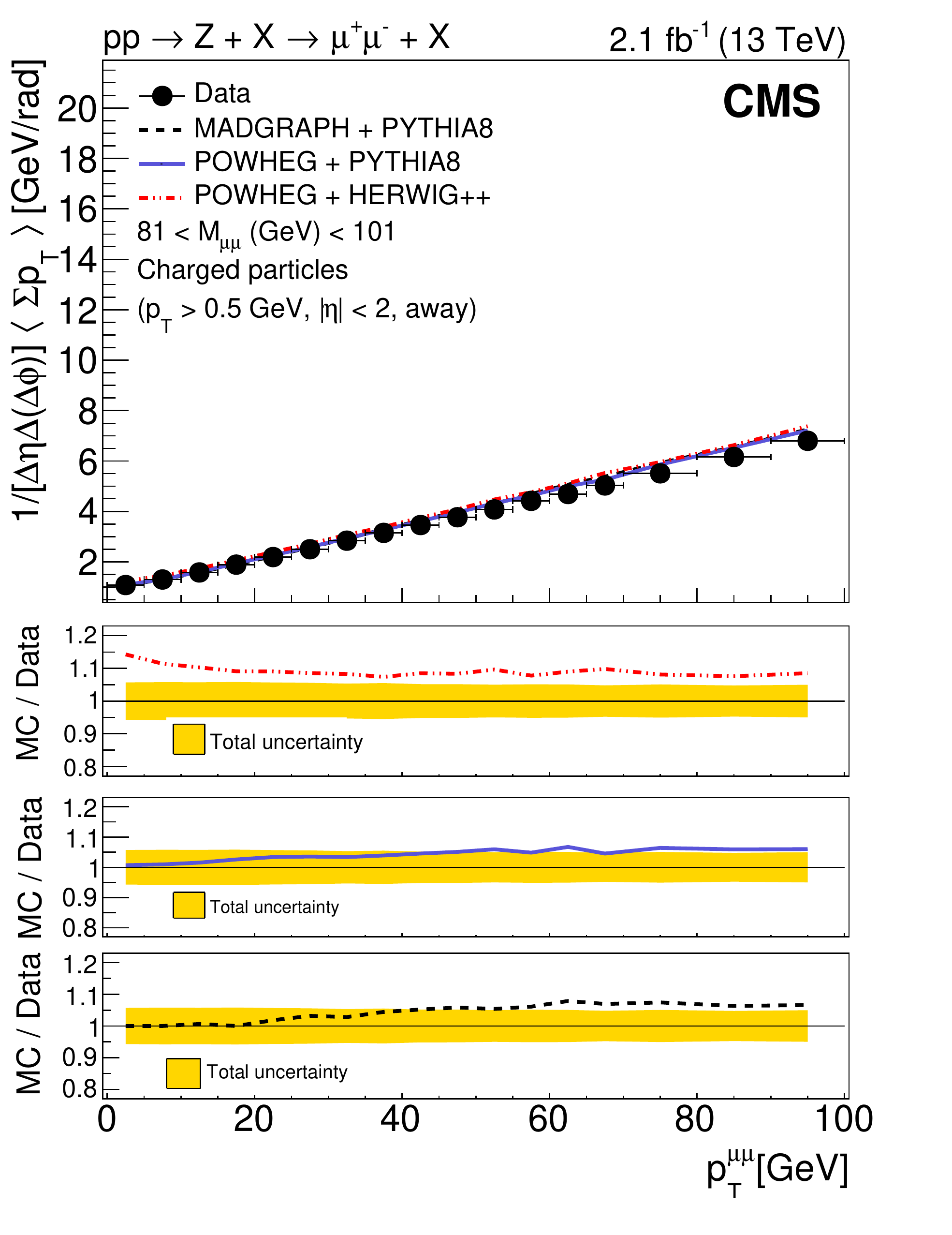}
\end{center}
\caption {{Unfolded distributions of particle density (left) and $\Sigma \pt$ density (right) in Z events in the \textit{away} region as a function of \pTmumu, compared to various model predictions: \MADGRAPH~+~\PYTHIA{8} (dashed line), \POWHEG~+~\PYTHIA{8} (solid line), and \POWHEG~+~\HERWIG{++} (dashed-dotted line). The bottom panels of each plot show the ratios of the simulations to the measured distributions. The bands in the bottom panels represent the statistical and systematic uncertainties added in quadrature.}}
\label{fig:dataMC13TeV_away}
\end{figure}

\begin{figure}[htbp]
\begin{center}
\includegraphics[width=0.49\textwidth]{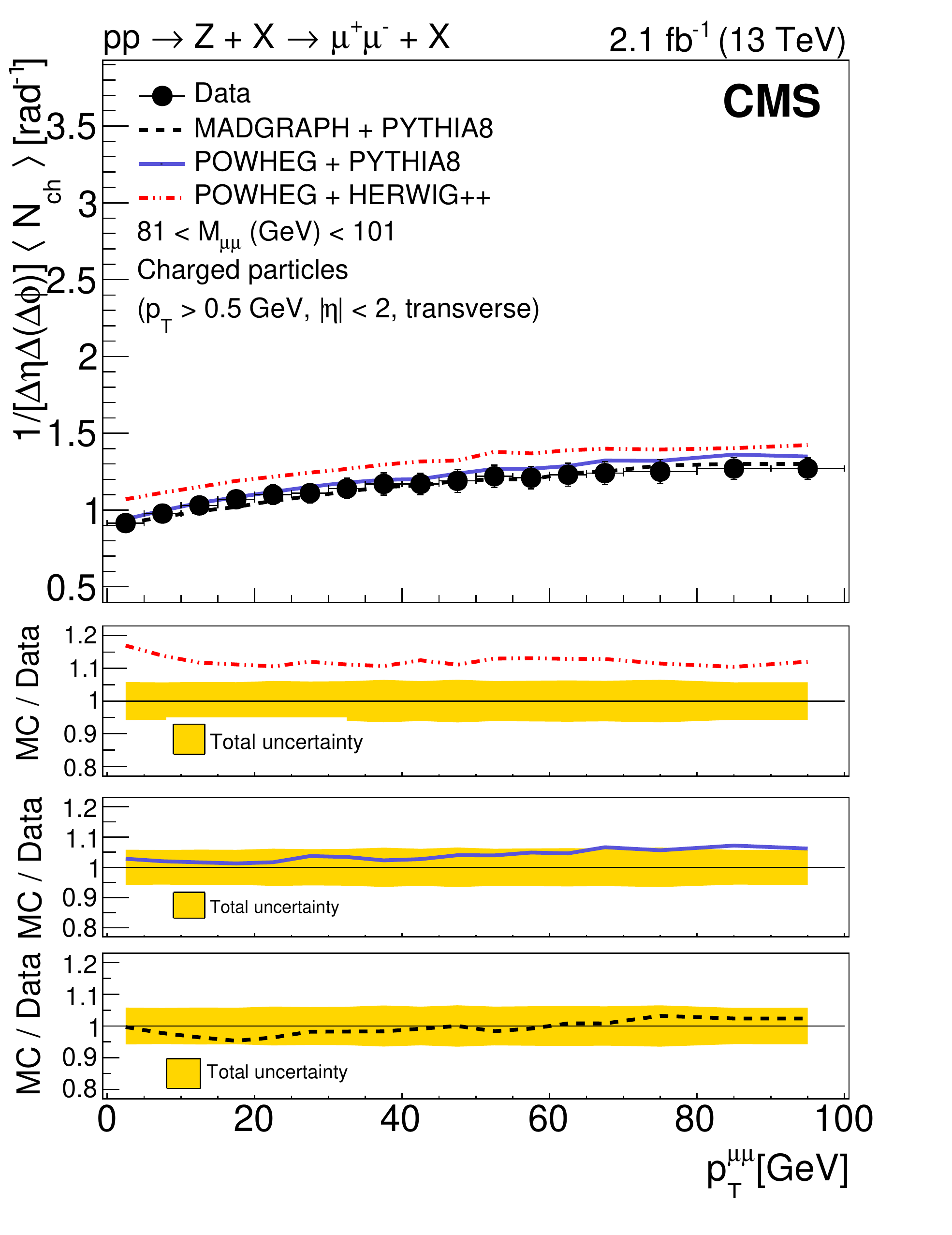} \hfil
\includegraphics[width=0.49\textwidth]{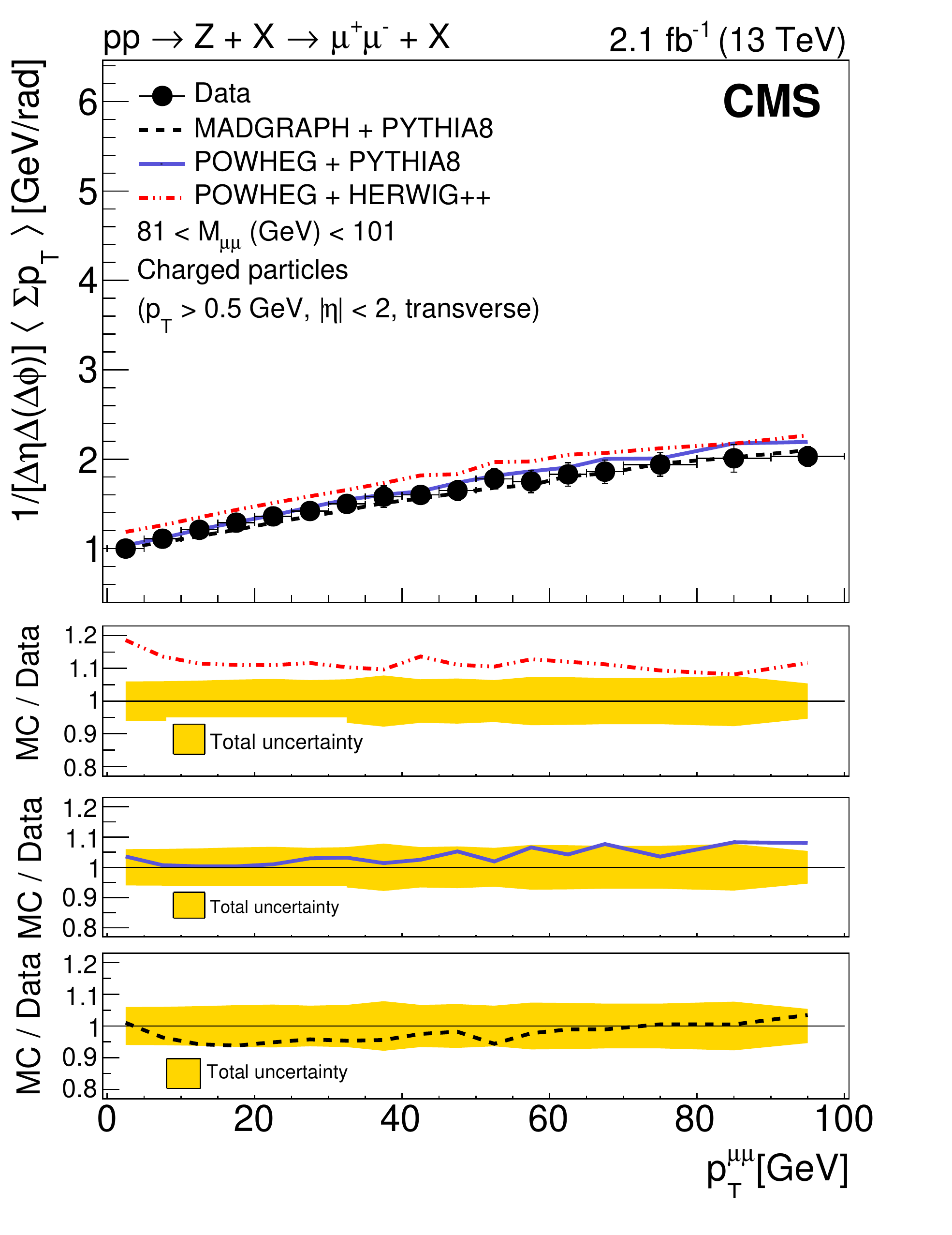}
\end{center}
\caption {{Unfolded distributions of particle density (left) and $\Sigma \pt$ density (right) in Z events in the \textit{transverse} region as a function of \pTmumu, compared to various model predictions: \MADGRAPH~+~\PYTHIA{8} (dashed line), \POWHEG~+~\PYTHIA{8} (solid line), and \POWHEG~+~\HERWIG{++} (dashed-dotted line). The bottom panels of each plot show the ratios of the simulations to the measured distributions. The bands in the bottom panels represent the statistical and systematic uncertainties added in quadrature.}}
\label{fig:dataMC13TeV_trans}
\end{figure}

\begin{figure}[htbp]
\begin{center}
\includegraphics[width=0.49\textwidth]{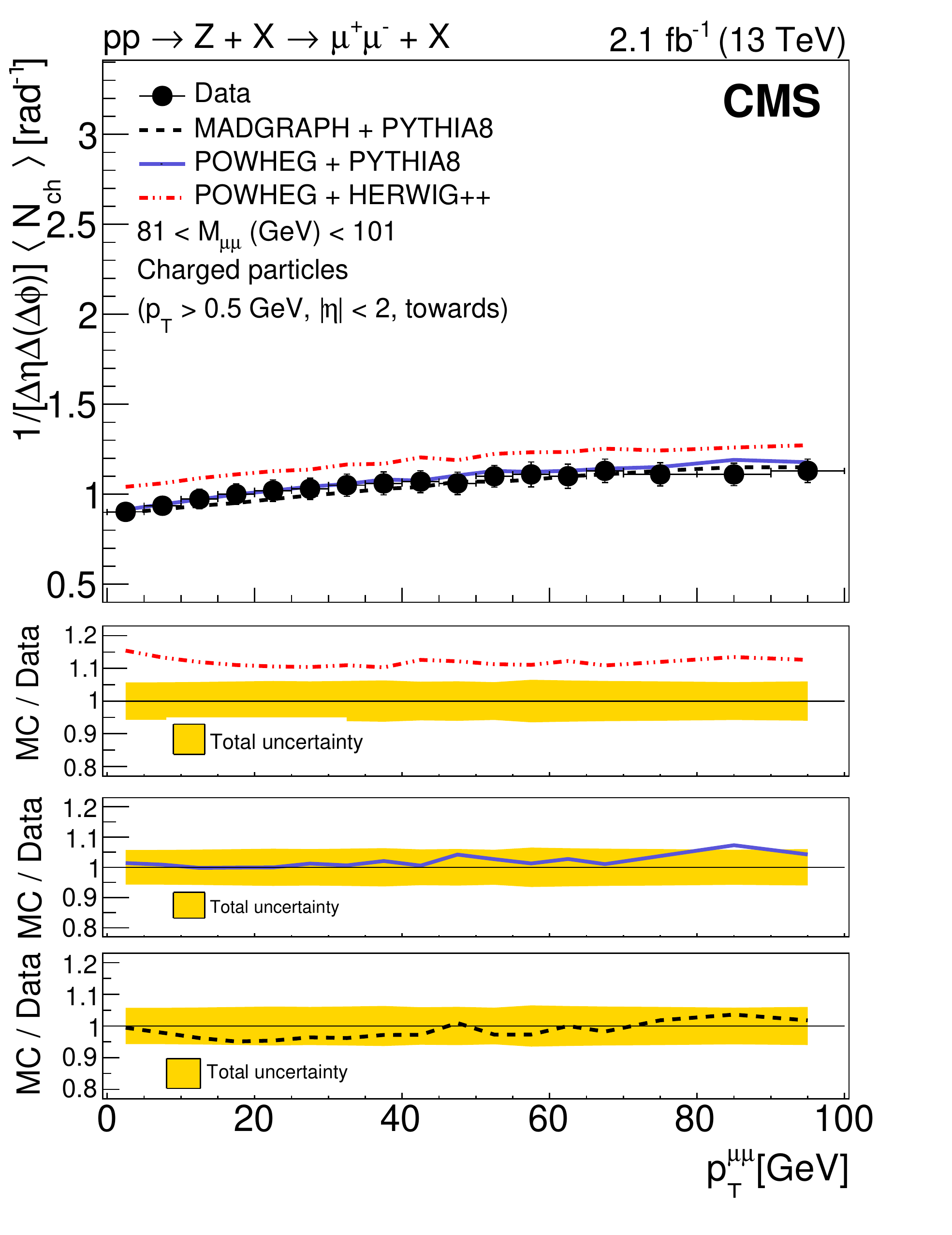} \hfil
\includegraphics[width=0.49\textwidth]{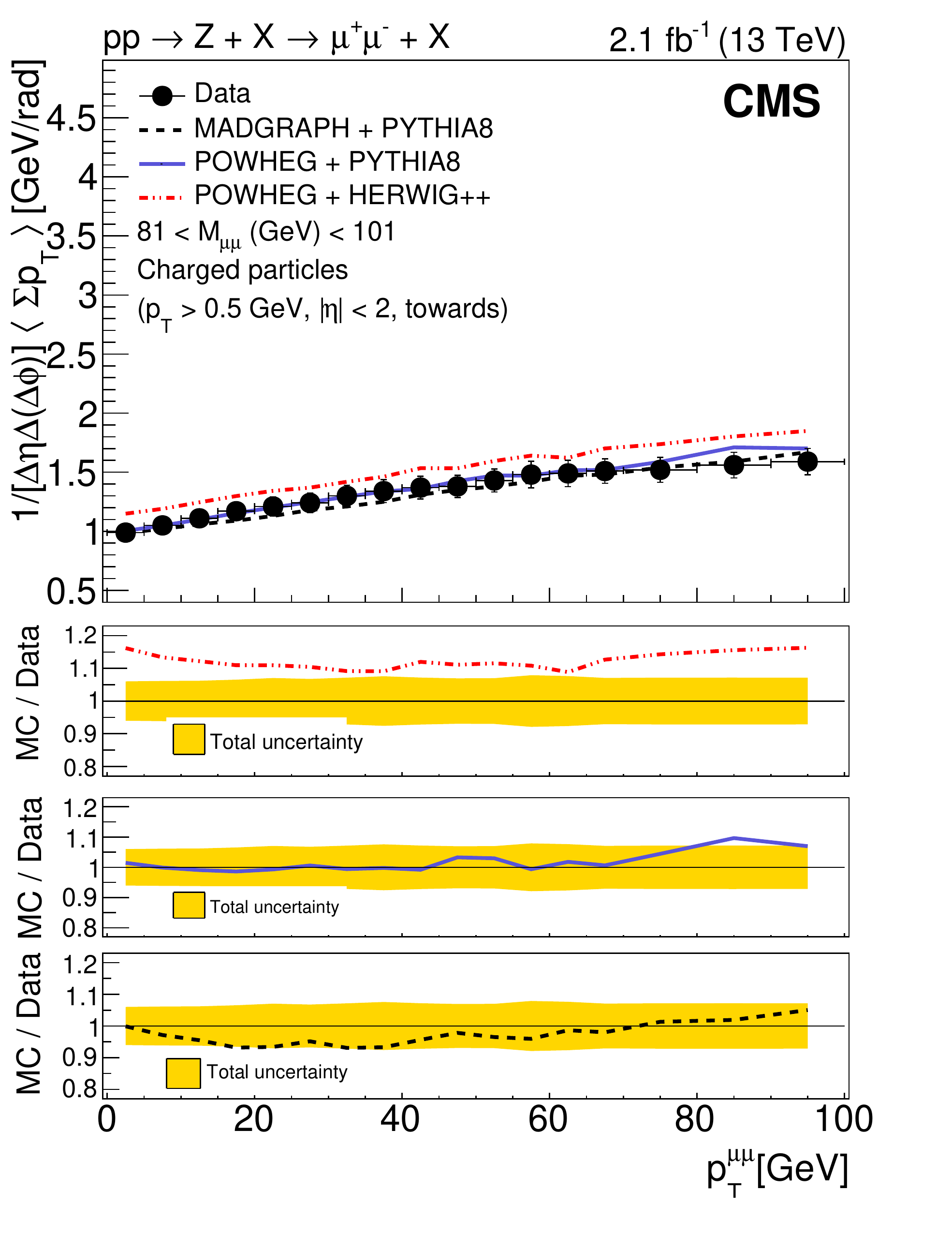}
\end{center}
\caption {{Unfolded distributions of particle density (left) and $\Sigma \pt$ density (right) in Z events in the \textit{towards} region as a function of \pTmumu, compared to various model predictions: \MADGRAPH~+~\PYTHIA{8} (dashed line), \POWHEG~+~\PYTHIA{8} (solid line), and \POWHEG~+~\HERWIG{++} (dashed-dotted line). The bottom panels of each plot show the ratios of the simulations to the measured distributions. The bands in the bottom panels represent the statistical and systematic uncertainties added in quadrature.}} \label{fig:dataMC13TeV_towards}
\end{figure}

Figures~\ref{fig:dataMC13TeV_away}--\ref{fig:dataMC13TeV_towards} present data-model comparisons of the UE distributions as a function of the Z boson \pt in the \textit{away}, \textit{transverse}, and \textit{towards} regions, respectively. The bottom panel of each plot presents the ratio of the simulated to the measured distributions. The \POWHEG sample, which uses \HERWIG{++} for parton showering and hadronization, overestimates the UE
activity by 10--15\% in all topological regions, whereas when \PYTHIA{8} is used the measured distributions are reproduced within 5\%. The \MADGRAPH sample in combination with \PYTHIA{8} also reproduces the measurement within 5\%. The \MCATNLO predictions (not shown in the figures) have the same level of agreement with the data as \MADGRAPH. Color reconnection between the produced partons influences the multiplicity and \pt of final-state particles. Its global impact in the measured UE observables is evaluated by comparing the \PYTHIA{8} predictions with and without CR, and is found to be negligible.

\begin{figure}[htbp]
\begin{center}
\includegraphics[width=0.90\textwidth]{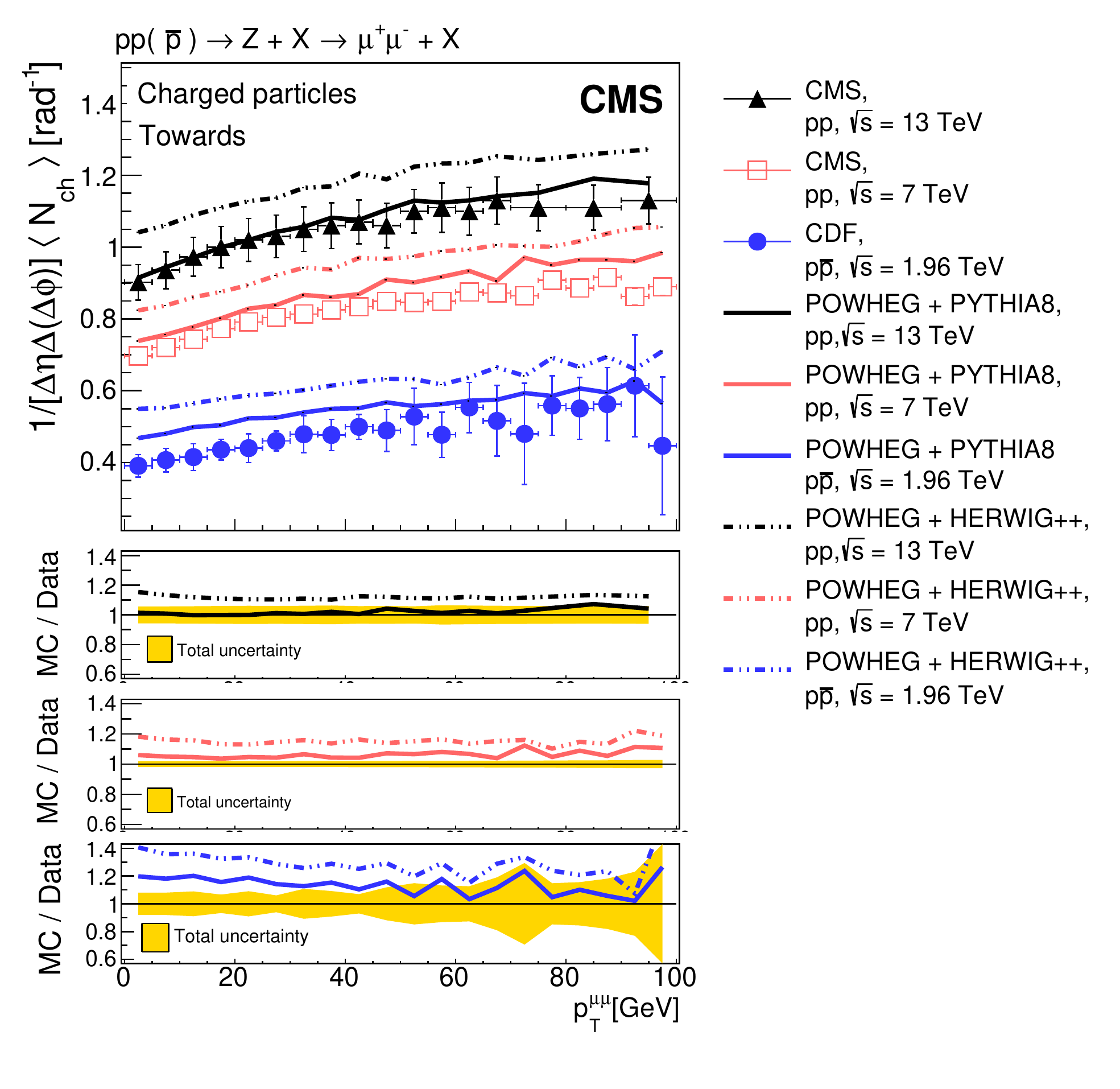}
\end{center}
\caption {{
Comparison of the particle density measured in Z events at $\sqrt{s} = 13\TeV$ with that at 7 (CMS)~\cite{uedycms} and 1.96\TeV (CDF)~\cite{uecdf} in the \textit{towards} region as a function of \pTmumu. The data are also compared with the
model predictions of \POWHEG~+~\PYTHIA{8} (solid line) and \POWHEG~+~\HERWIG{++} (dashed-dotted line). The bottom panels of each plot show the ratios of the model predictions to the measurements. The bands in the bottom panels represent
the statistical and systematic uncertainties added in quadrature.}} \label{fig:allEnergyComp}
\end{figure}

\begin{figure}[htbp]
\begin{center}
\includegraphics[width=0.9\textwidth]{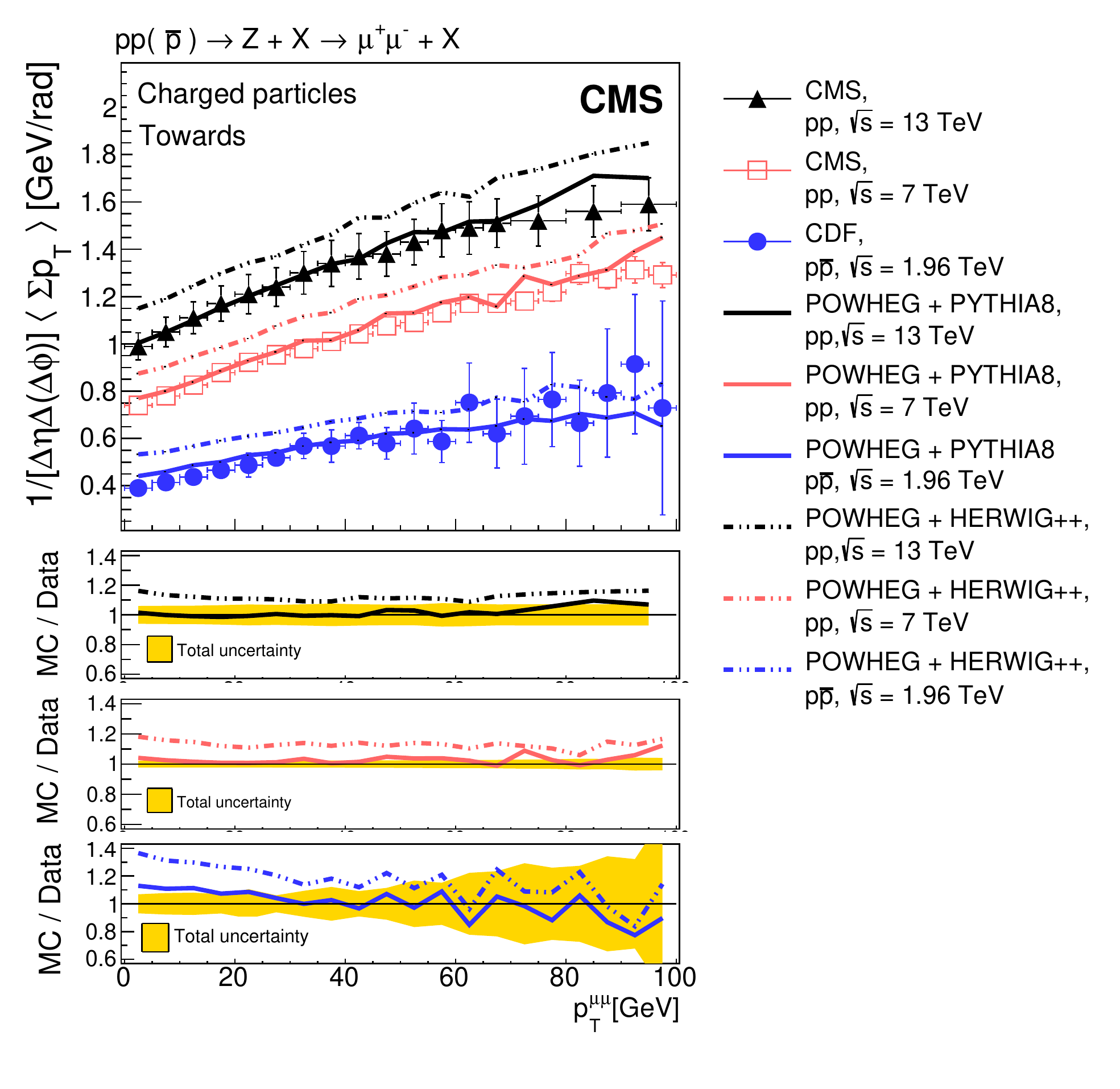}
\end{center}
\caption {{
Comparison of the $\Sigma \pt$ density measured in Z events at $\sqrt{s} = 13\TeV$ with that at 7 (CMS)~\cite{uedycms} and 1.96\TeV (CDF)~\cite{uecdf} in the \textit{towards} region as a function of \pTmumu. The data are also compared with the model predictions of
\POWHEG~+~\PYTHIA{8} (solid line) and \POWHEG~+~\HERWIG{++} (dashed-dotted line). The bottom panels of each plot show
the ratios of the model predictions to the measurements. The bands in the bottom panels represent the statistical and systematic uncertainties added in quadrature.}} \label{fig:allEnergyComp1}
\end{figure}

\begin{figure}[htbp]
\begin{center}
\includegraphics[width=0.9\textwidth]{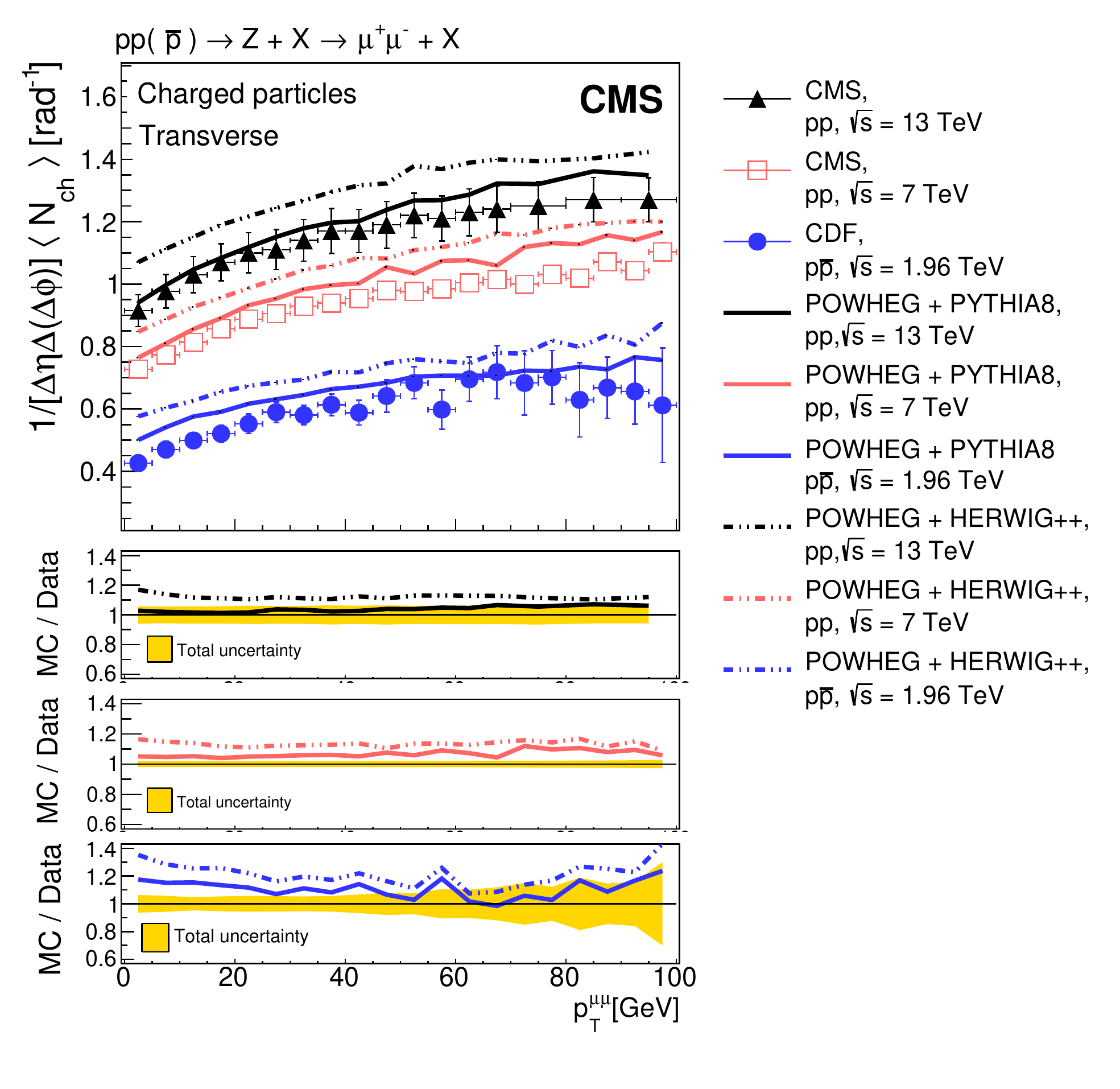}
\end{center}
\caption {{Comparison of the particle density measured in Z events at $\sqrt{s} = 13\TeV$ with that at 7 (CMS)~\cite{uedycms} and 1.96\TeV (CDF)~\cite{uecdf} in the \textit{transverse} region as a function of \pTmumu. The data are also compared with the model predictions
of \POWHEG~+~\PYTHIA{8} (solid line) and \POWHEG~+~\HERWIG{++} (dashed-dotted line). The bottom panels of each plot show the
ratios of model predictions to the measurements. The bands in the bottom panels represent the statistical and systematic uncertainties added in quadrature.}} \label{fig:allEnergyComp2}
\end{figure}

\begin{figure}[htbp]
\begin{center}
\includegraphics[width=0.9\textwidth]{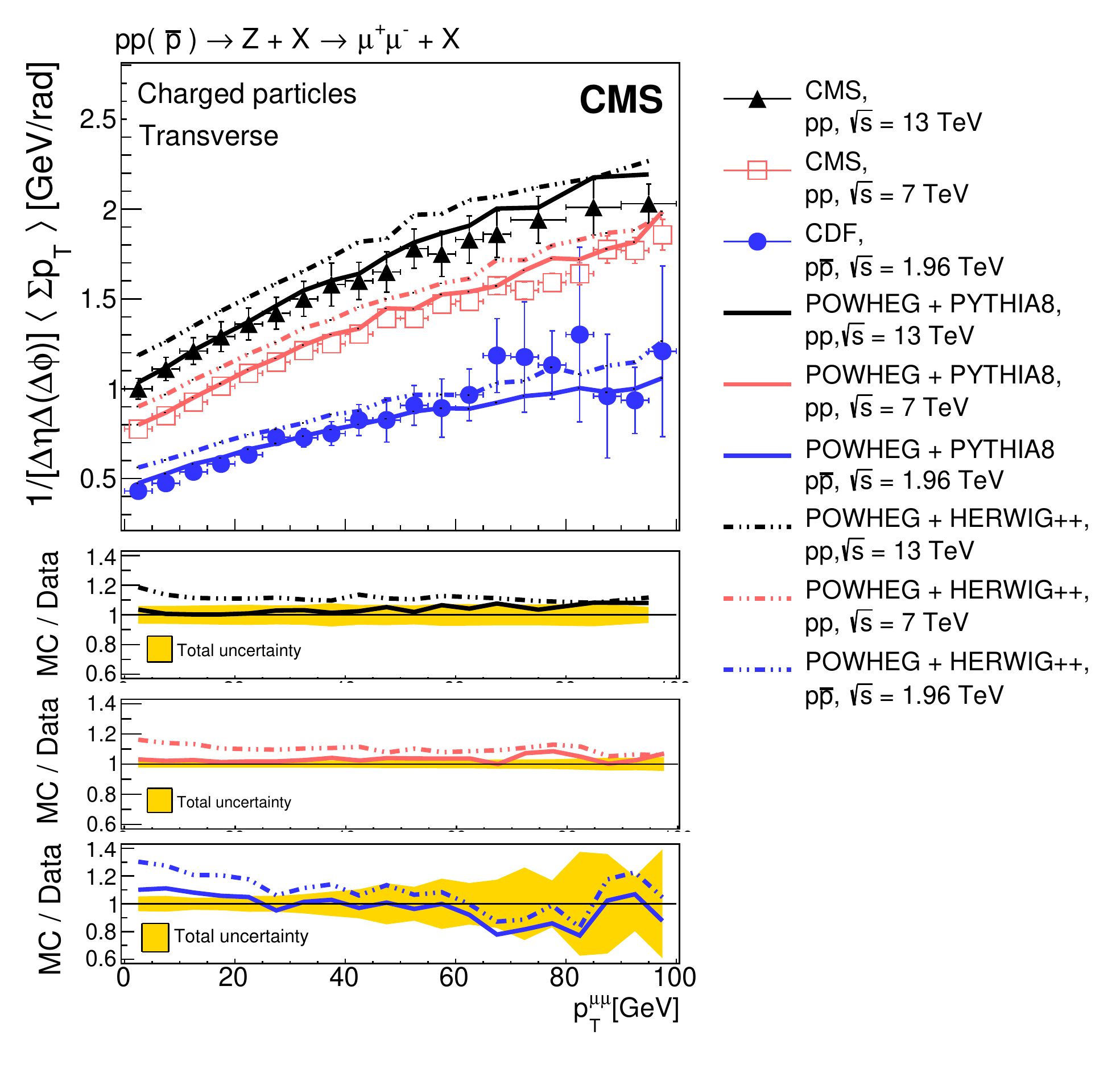}
\end{center}
\caption {{Comparison of the $\Sigma \pt$ density measured in Z events at $\sqrt{s} = 13\TeV$ with that at 7 (CMS)~\cite{uedycms} and 1.96\TeV (CDF)~\cite{uecdf} in the \textit{transverse} region as a function of \pTmumu. The data are also compared with the predictions of \POWHEG~+~\PYTHIA{8} (solid line) and \POWHEG~+~\HERWIG{++} (dashed-dotted line). The bottom panels of each plot show the ratios of the model predictions to the measurements. The bands in the bottom panels represent the statistical and systematic uncertainties added in quadrature.}} \label{fig:allEnergyComp3}
\end{figure}

To understand the evolution of the UE activity with $\sqrt{s}$, the present measurement is compared
with results obtained at $\sqrt{s} = 1.96\TeV$ at the Tevatron and at 7\TeV at the LHC. As the \textit{away}
region is dominated by the jet balancing the Z boson, the particle activity in this region is not considered for this
specific study. Figures~\ref{fig:allEnergyComp}--\ref{fig:allEnergyComp3} show the
UE activity as a function of \pTmumu at $\sqrt{s}= 1.96$, 7, and 13\TeV.
The predictions of \POWHEG with \PYTHIA{8} as well as with \HERWIG{++} are also shown. The
ratios of the simulations to the measurements are plotted in the bottom panel of each plot. The
\POWHEG~+~\PYTHIA{8} predictions reproduce the measurements within 10\% at $\sqrt{s}$
of 1.96 \TeV and 7 \TeV, and within 5\% at 13 \TeV. The combination of \POWHEG and \HERWIG{++}
describes the measurements within 10--15, 10--20, and 20--40\% at $\sqrt{s}$ of 1.96, 7,
and 13 \TeV, respectively.

\begin{figure}[htbp]
\begin{center}
\includegraphics[width=0.49\textwidth]{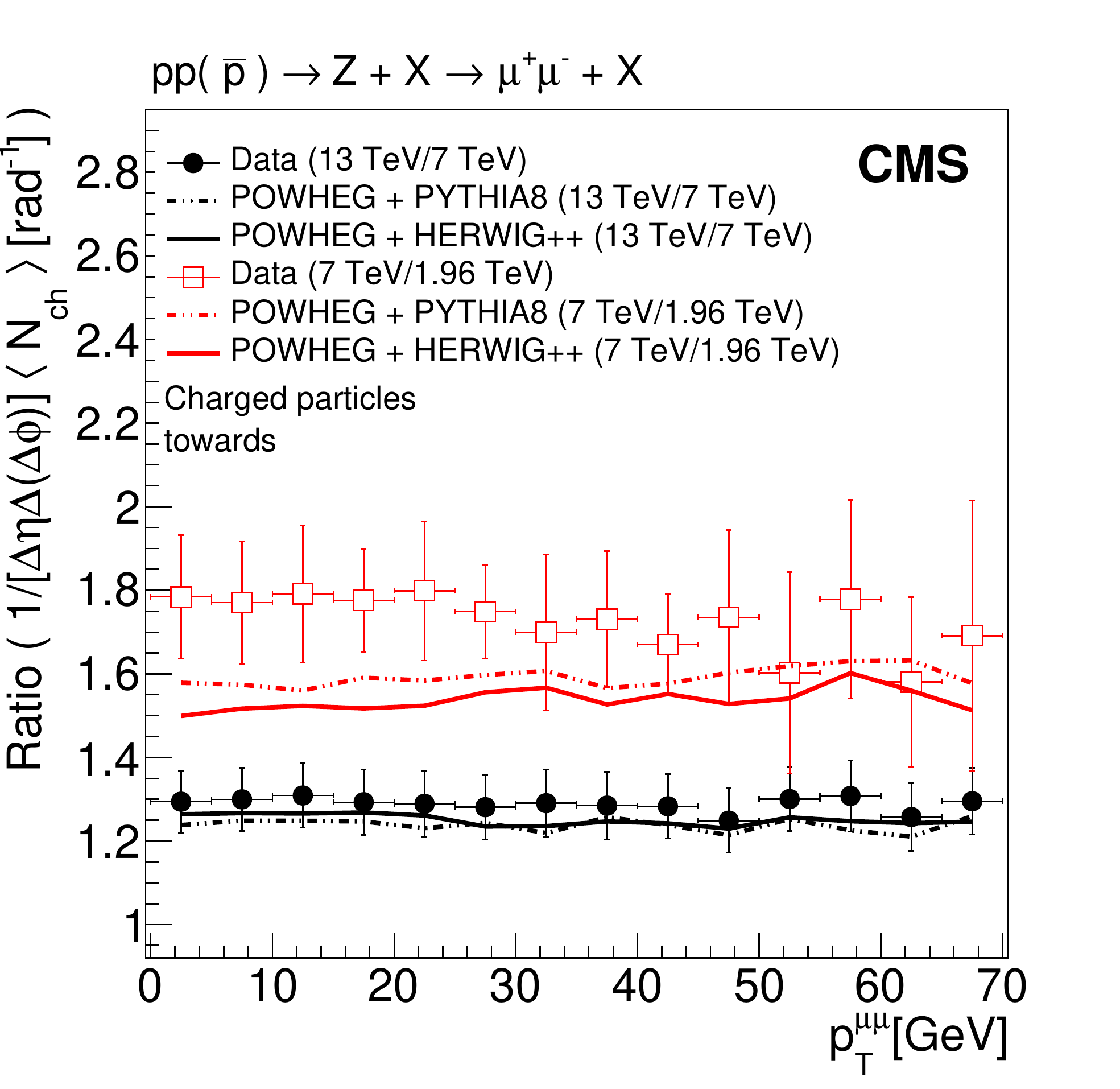} \hfil
\includegraphics[width=0.49\textwidth]{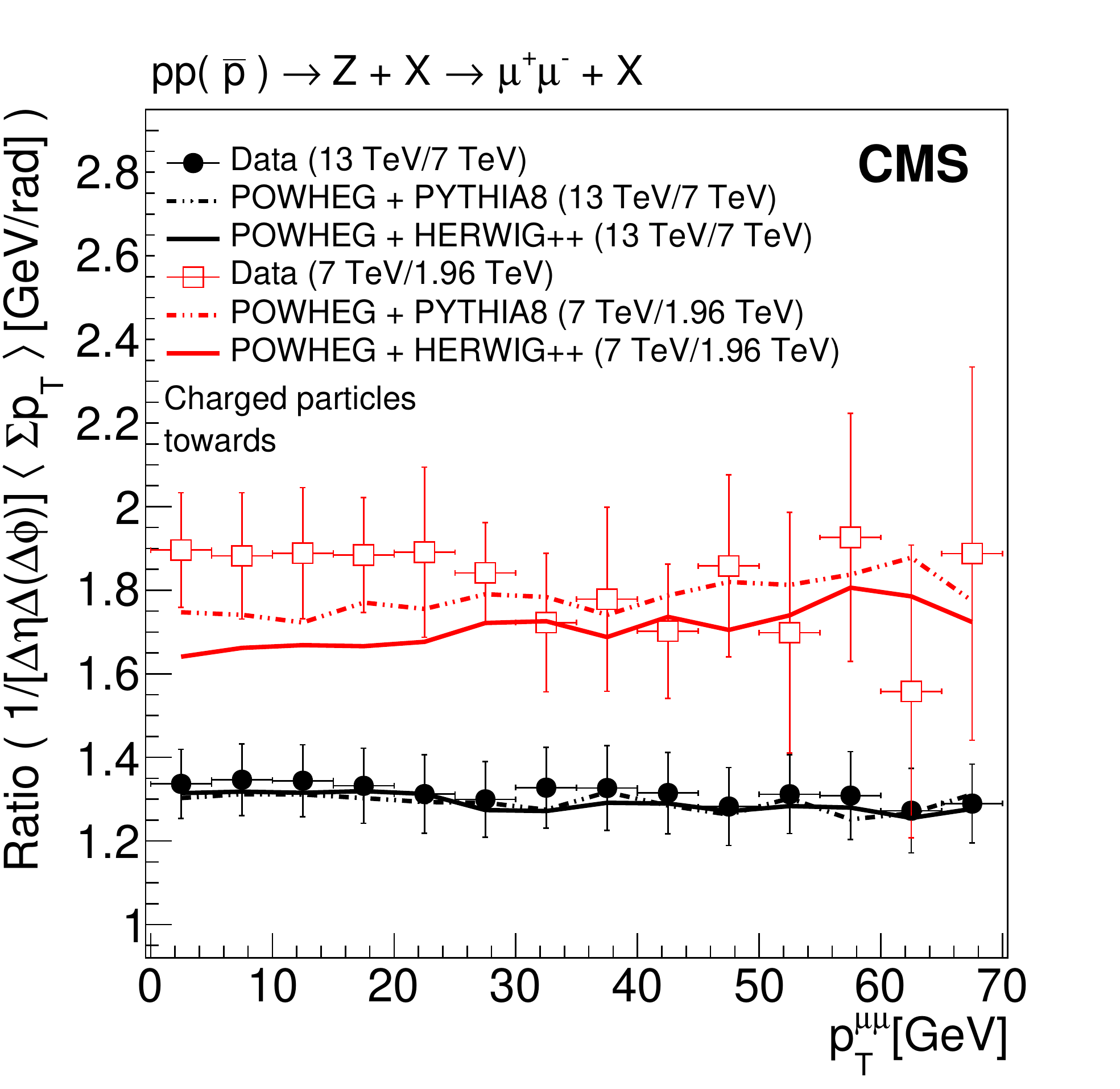} \\
\includegraphics[width=0.49\textwidth]{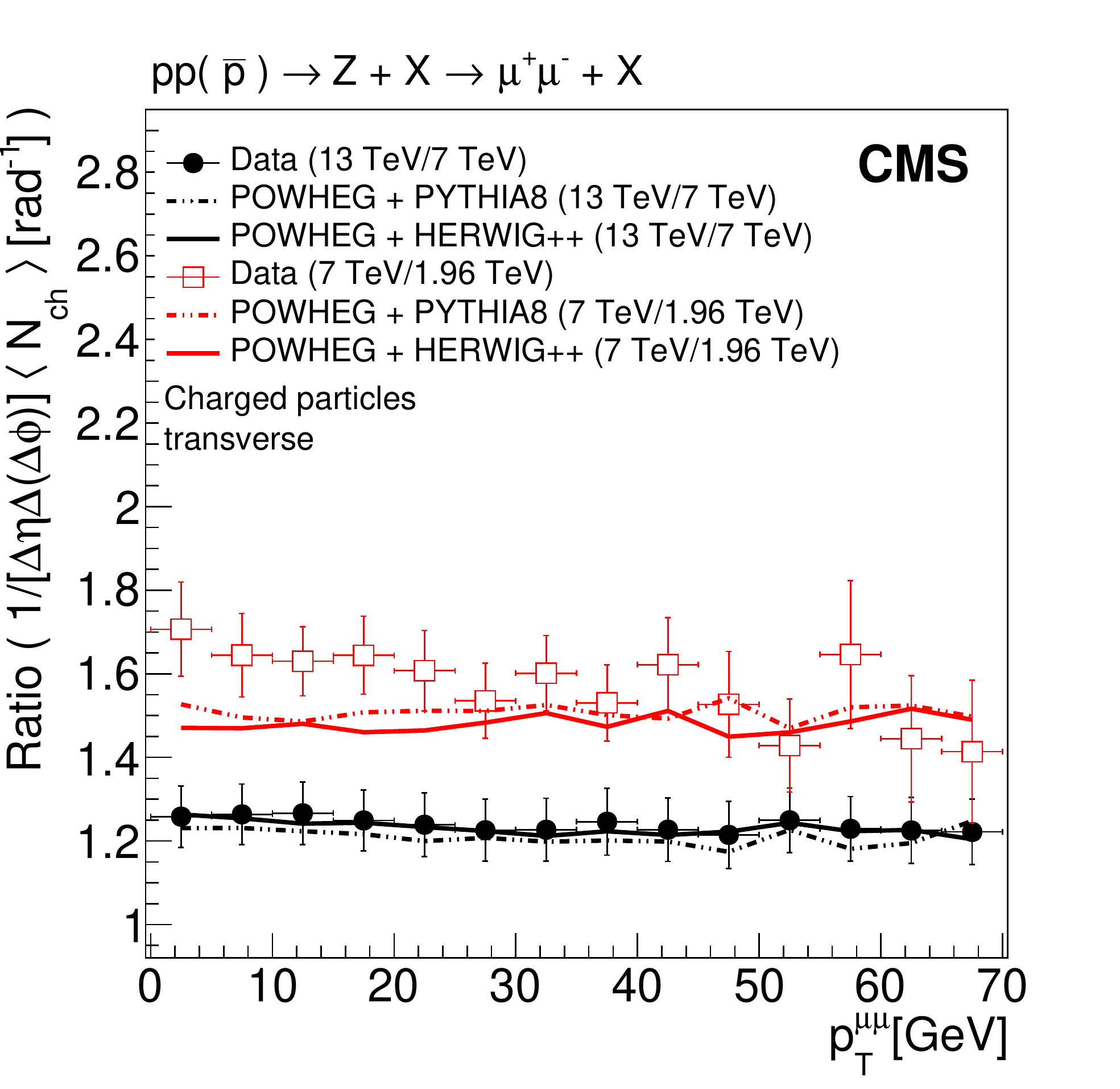} \hfil
\includegraphics[width=0.49\textwidth]{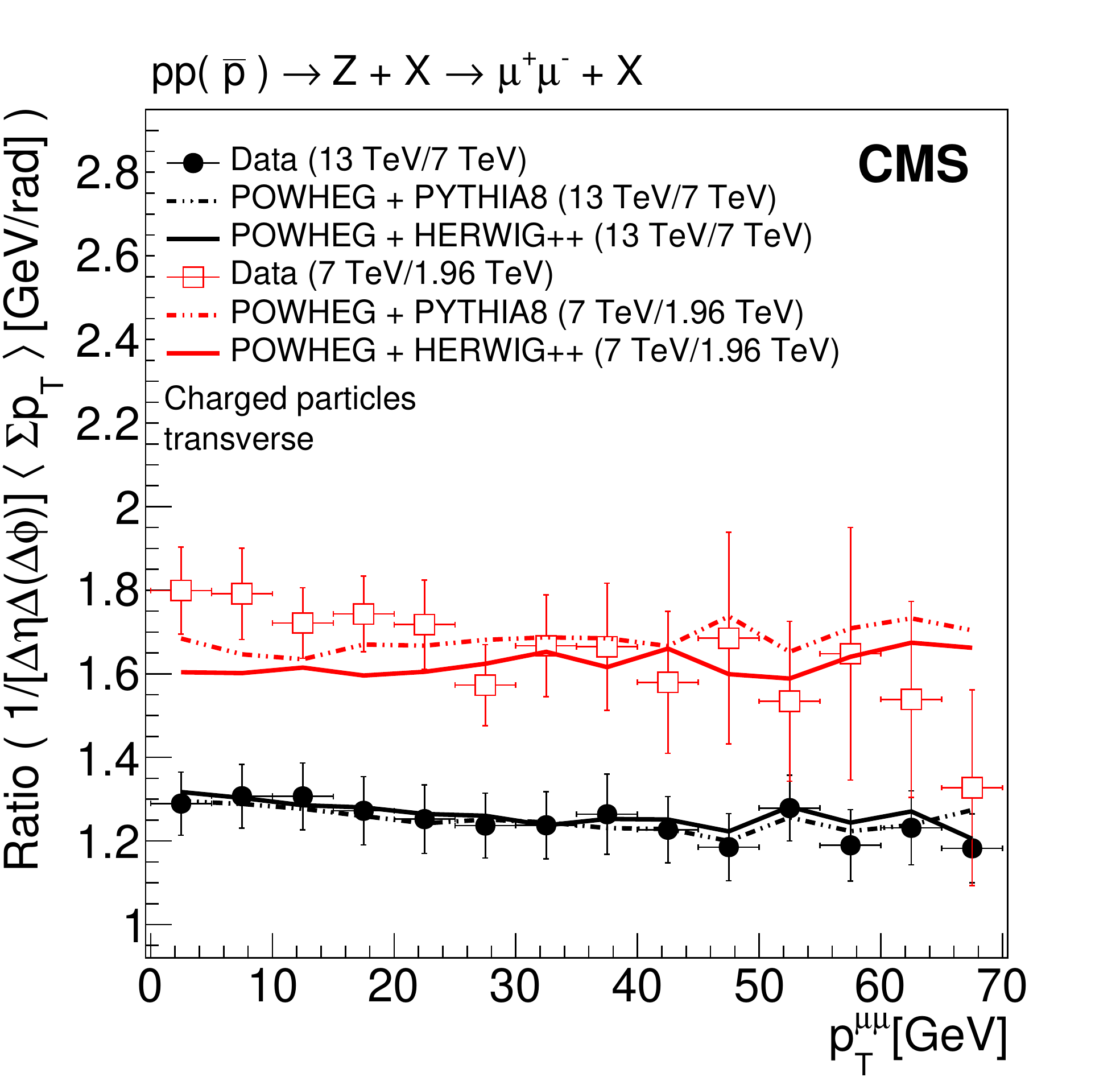}
\end{center}
\caption {{Comparison of the increase in UE activity in Z events, from $\sqrt{s} = 1.96 \TeV$ (CDF)~\cite{uecdf} to 7\TeV (CMS)~\cite{uedycms}, with that from $\sqrt{s} = 7\TeV$ (CMS) to 13\TeV (CMS) in the \textit{towards} (top) and \textit{transverse} (bottom) regions. Panels on the left show the particle density, whereas panels on the right show the $\Sigma \pt$ density as a function of \pTmumu . The data distributions are
also compared with predictions of \POWHEG~+~\PYTHIA{8} (dashed-dotted line) and \POWHEG~+~\HERWIG{++} (solid line).
The error bars represent the statistical and systematic uncertainties added in quadrature.
}}
\label{fig:allEnergyComp4}
\end{figure}

The data show a significant increase in the UE activity with $\sqrt{s}$, which is qualitatively described by the model predictions. The collision energy evolution is quantified in Fig.~\ref{fig:allEnergyComp4},
which shows the ratio of the UE activities at 13 and 7 \TeV, and at 1.96 and 7 \TeV, for the data and the simulations. An increase of 25--30\% in particle and $\Sigma \pt$ densities is observed as the collision energy increases from 7 to 13\TeV.
This behavior is quantitatively well described by \POWHEG~+~\PYTHIA{8} and
\POWHEG~+~\HERWIG{++}. As the collision energy increases from 1.96 to 7\TeV, the UE
activity increases by 60--80\% for both the particle and $\Sigma \pt$ densities. Event generators predict a
slower rise, but the agreement improves at higher values of \pTmumu. The increase in particle and $\Sigma \pt$
densities from 7 to 13 \TeV is consistent with that observed in the leading jet/track analyses~\cite{uedycms,ueatlas2}.

\begin{figure}[htbp]
\begin{center}
\includegraphics[width=0.49\textwidth]{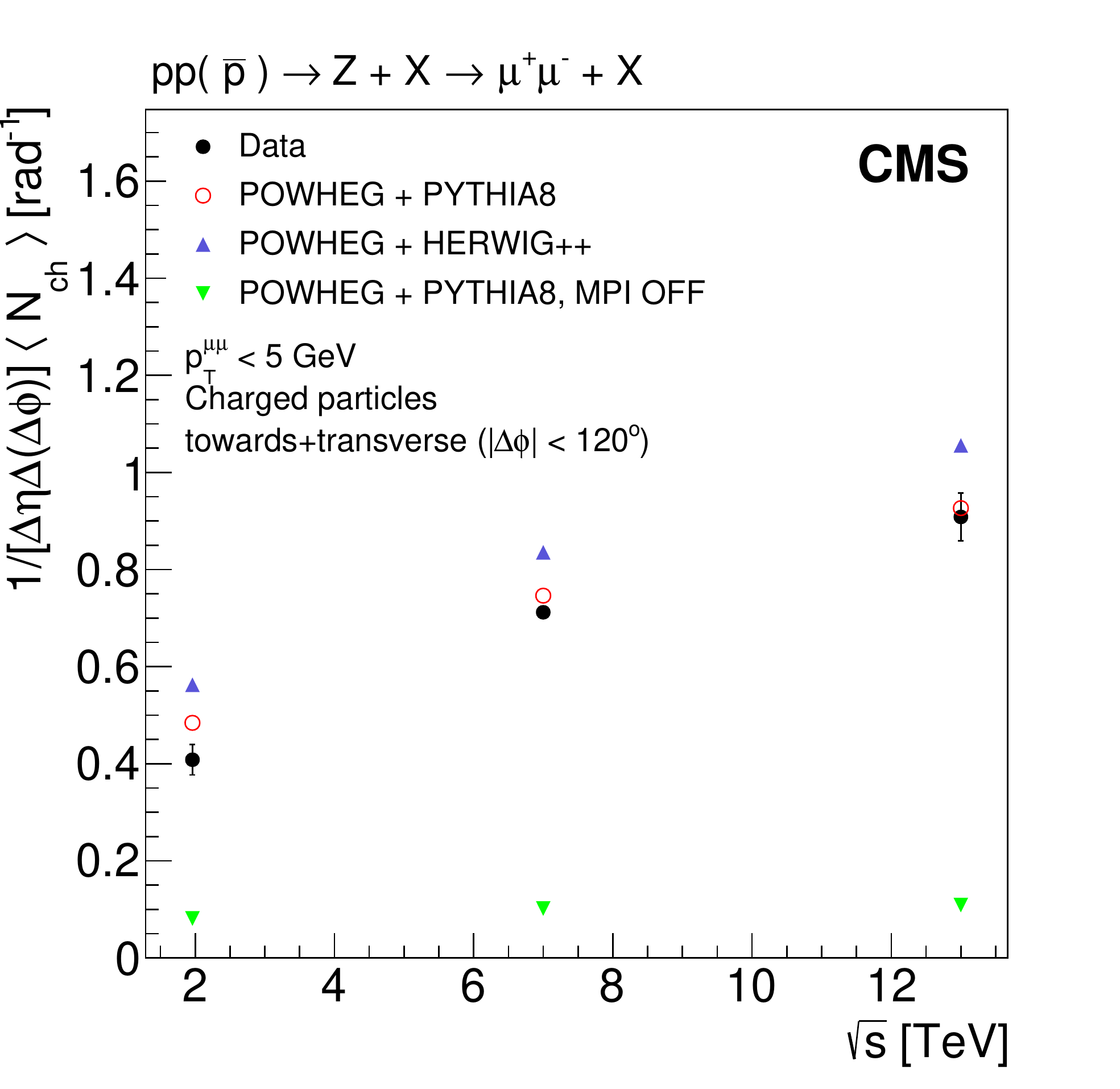} \hfil
\includegraphics[width=0.49\textwidth]{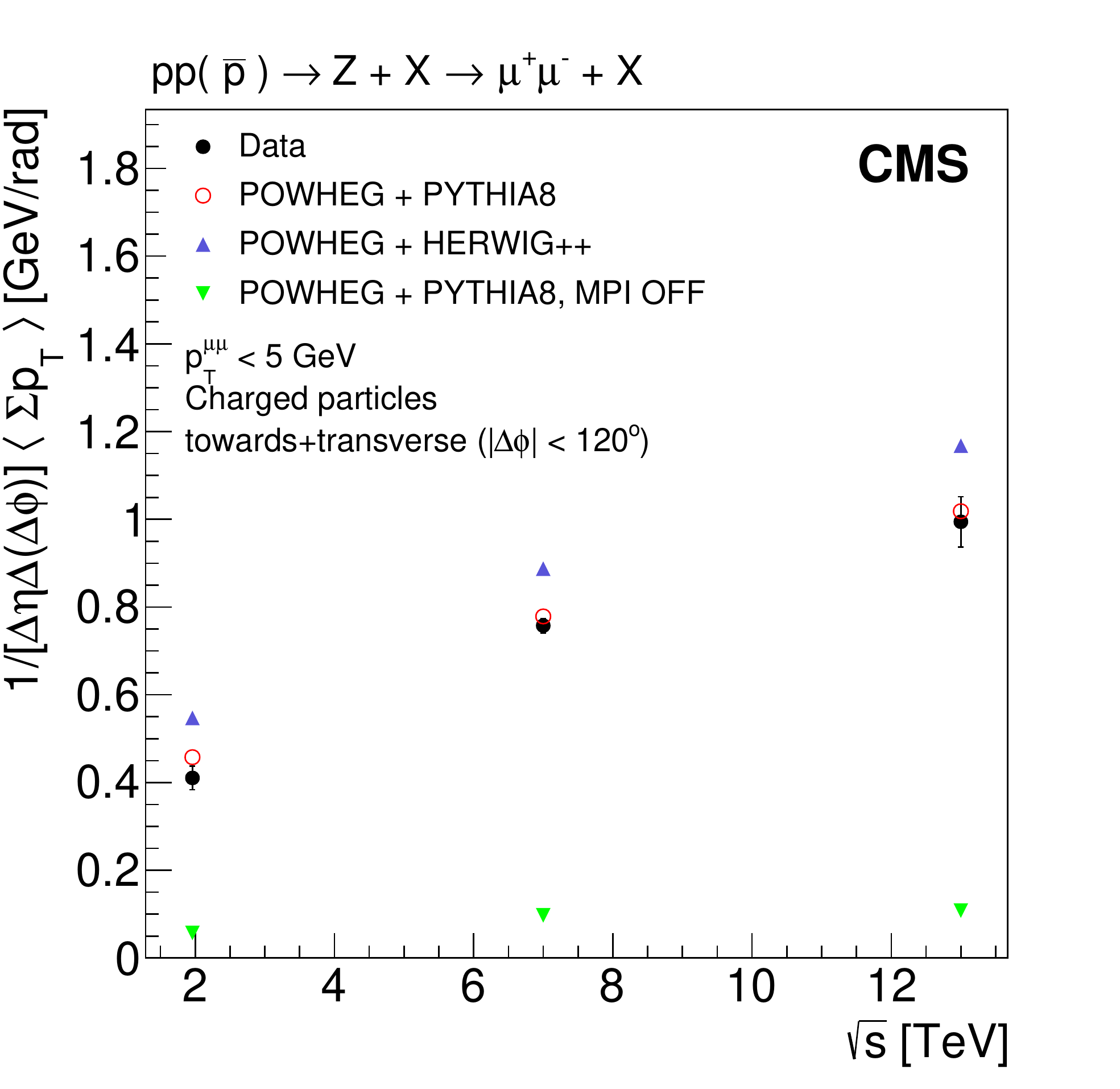}
\end{center}
\caption {{Average particle density (left) and average $\Sigma \pt$ density (right) for Z events with $ \pTmumu< 5\GeV$ as a function of the center-of-mass energy, measured by CMS and CDF~\cite{uecdf} in the combined \textit{towards} + \textit{transverse} regions, compared to predictions from \POWHEG~+~\PYTHIA{8}, \POWHEG~+~\HERWIG{++}, and \POWHEG~+~\PYTHIA{8} without MPI. The error bars represent the statistical and systematic uncertainties added in quadrature.}} \label{fig:dataMCpickPt}
\end{figure}

To further quantify the energy dependence of the UE activity, events with a \pTmumu smaller than 5\GeV are studied.
Setting an upper limit on \pTmumu reduces the ISR and FSR contributions and the remaining UE activity stems mainly from MPI. With the
requirement $\pTmumu < 5\GeV$, the UE activity is similar in the \textit{towards} and \textit{transverse} regions.
Therefore, the UE activity is combined in these two regions. Figure~\ref{fig:dataMCpickPt} shows the UE activity, with the \pTmumu $<$ 5\GeV requirement, as a function of $\sqrt{s}$ for data compared to model predictions. There is a significant increase, by a factor 2--2.5, as the collision energy rises from 1.96 to 13\TeV, which is qualitatively reproduced by \POWHEG. The energy evolution is better described by \POWHEG with \PYTHIA{8}, whereas hadronization with \HERWIG{++} overestimates the UE activity at all collision energies. The comparison of the distributions with and without MPI indicates that the ISR and FSR contributions, which increase slowly with center-of-mass energy, are small.

The CUETP8M1 and EE5C tunes employed here are mostly obtained from fits to minimum-bias measurements and UE measurements with leading
jets or leading tracks. The fact that these tunes reproduce globally well the present data supports the
hypothesis that the UE activity is independent of the hard process. The present study also confirms that the
collision energy dependence of the UE activity is similar for different hard processes. Unlike UE studies with a leading track/jet, the present measurements provide new handles to better understand the evolution of ISR, FSR, and MPI contributions separately, as functions of the event energy scale and the collision energy.

\section{Summary}
\label{sec:summary}

This paper presents a measurement of the underlying event (UE) activity using inclusive Z boson production events in proton-proton collisions at a center-of-mass energy of 13\TeV. The data correspond to an integrated luminosity of 2.1\fbinv. The
UE activity, quantified in terms of charged particle and $\Sigma \pt$ densities, is measured as a function
of the \pt of the muon pair from the Z boson decay. The distributions are corrected for detector
effects and compared to various model predictions. The \MADGRAPH and \POWHEG generators, with parton showering and hadronization modeled with \PYTHIA{8} using the CUET8PM1 tune, reproduce the measurements within 5\%.
The combination of \POWHEG and \HERWIG{++} (tune EE5C) overestimates the measurements by 10--15\%. The
present results are also compared with previous measurements at 1.96 and 7\TeV. The UE activity
almost doubles as the collision energy increases from 1.96 to 13\TeV.
Monte Carlo event generators provide a reasonable description of the evolution of the UE activity as
the collision energy rises from 1.96 to 13\TeV, although they tend to underestimate its increase in
the 1.96--7\TeV range. The overall good description of the UE activity in Z boson events by Monte Carlo generators previously tuned
to minimum-bias and leading track/jet UE measurements confirms the universality of the physical
processes producing the underlying event in pp collisions at high energies.

\newpage

\begin{acknowledgments}
\hyphenation{Bundes-ministerium Forschungs-gemeinschaft Forschungs-zentren Rachada-pisek} We congratulate our colleagues in the CERN accelerator departments for the excellent performance of the LHC and thank the technical and administrative staffs at CERN and at other CMS institutes for their contributions to the success of the CMS effort. In addition, we gratefully acknowledge the computing centers and personnel of the Worldwide LHC Computing Grid for delivering so effectively the computing infrastructure essential to our analyses. Finally, we acknowledge the enduring support for the construction and operation of the LHC and the CMS detector provided by the following funding agencies: the Austrian Federal Ministry of Science, Research and Economy and the Austrian Science Fund; the Belgian Fonds de la Recherche Scientifique, and Fonds voor Wetenschappelijk Onderzoek; the Brazilian Funding Agencies (CNPq, CAPES, FAPERJ, and FAPESP); the Bulgarian Ministry of Education and Science; CERN; the Chinese Academy of Sciences, Ministry of Science and Technology, and National Natural Science Foundation of China; the Colombian Funding Agency (COLCIENCIAS); the Croatian Ministry of Science, Education and Sport, and the Croatian Science Foundation; the Research Promotion Foundation, Cyprus; the Secretariat for Higher Education, Science, Technology and Innovation, Ecuador; the Ministry of Education and Research, Estonian Research Council via IUT23-4 and IUT23-6 and European Regional Development Fund, Estonia; the Academy of Finland, Finnish Ministry of Education and Culture, and Helsinki Institute of Physics; the Institut National de Physique Nucl\'eaire et de Physique des Particules~/~CNRS, and Commissariat \`a l'\'Energie Atomique et aux \'Energies Alternatives~/~CEA, France; the Bundesministerium f\"ur Bildung und Forschung, Deutsche Forschungsgemeinschaft, and Helmholtz-Gemeinschaft Deutscher Forschungszentren, Germany; the General Secretariat for Research and Technology, Greece; the National Scientific Research Foundation, and National Innovation Office, Hungary; the Department of Atomic Energy and the Department of Science and Technology, India; the Institute for Studies in Theoretical Physics and Mathematics, Iran; the Science Foundation, Ireland; the Istituto Nazionale di Fisica Nucleare, Italy; the Ministry of Science, ICT and Future Planning, and National Research Foundation (NRF), Republic of Korea; the Lithuanian Academy of Sciences; the Ministry of Education, and University of Malaya (Malaysia); the Mexican Funding Agencies (BUAP, CINVESTAV, CONACYT, LNS, SEP, and UASLP-FAI); the Ministry of Business, Innovation and Employment, New Zealand; the Pakistan Atomic Energy Commission; the Ministry of Science and Higher Education and the National Science Centre, Poland; the Funda\c{c}\~ao para a Ci\^encia e a Tecnologia, Portugal; JINR, Dubna; the Ministry of Education and Science of the Russian Federation, the Federal Agency of Atomic Energy of the Russian Federation, Russian Academy of Sciences, the Russian Foundation for Basic Research and the Russian Competitiveness Program of NRNU ``MEPhI"; the Ministry of Education, Science and Technological Development of Serbia; the Secretar\'{\i}a de Estado de Investigaci\'on, Desarrollo e Innovaci\'on, Programa Consolider-Ingenio 2010, Plan de Ciencia, Tecnolog\'{i}a e Innovaci\'on 2013-2017 del Principado de Asturias and Fondo Europeo de Desarrollo Regional, Spain; the Swiss Funding Agencies (ETH Board, ETH Zurich, PSI, SNF, UniZH, Canton Zurich, and SER); the Ministry of Science and Technology, Taipei; the Thailand Center of Excellence in Physics, the Institute for the Promotion of Teaching Science and Technology of Thailand, Special Task Force for Activating Research and the National Science and Technology Development Agency of Thailand; the Scientific and Technical Research Council of Turkey, and Turkish Atomic Energy Authority; the National Academy of Sciences of Ukraine, and State Fund for Fundamental Researches, Ukraine; the Science and Technology Facilities Council, UK; the US Department of Energy, and the US National Science Foundation.

Individuals have received support from the Marie-Curie program and the European Research Council and Horizon 2020 Grant, contract No. 675440 (European Union); the Leventis Foundation; the A. P. Sloan Foundation; the Alexander von Humboldt Foundation; the Belgian Federal Science Policy Office; the Fonds pour la Formation \`a la Recherche dans l'Industrie et dans l'Agriculture (FRIA-Belgium); the Agentschap voor Innovatie door Wetenschap en Technologie (IWT-Belgium); the Ministry of Education, Youth and Sports (MEYS) of the Czech Republic; the Council of Scientific and Industrial Research, India; the HOMING PLUS program of the Foundation for Polish Science, cofinanced from European Union, Regional Development Fund, the Mobility Plus program of the Ministry of Science and Higher Education, the National Science Center (Poland), contracts Harmonia 2014/14/M/ST2/00428, Opus 2014/13/B/ST2/02543, 2014/15/B/ST2/03998, and 2015/19/B/ST2/02861, Sonata-bis 2012/07/E/ST2/01406; the National Priorities Research Program by Qatar National Research Fund; the Programa Severo Ochoa del Principado de Asturias; the Thalis and Aristeia programs cofinanced by EU-ESF and the Greek NSRF; the Rachadapisek Sompot Fund for Postdoctoral Fellowship, Chulalongkorn University and the Chulalongkorn Academic into Its 2nd Century Project Advancement Project (Thailand); the Welch Foundation, contract C-1845; and the Weston Havens Foundation (USA).

\end{acknowledgments}

\newpage

\bibliography{auto_generated}

\providecommand{\href}[2]{#2}\begingroup\raggedright\begin{thebibliography}{10}%
\makeatletter
\providecommand{\hrefCMSnoop }[0]{\@secondoftwo}%
\makeatother
\providecommand{\doi}{\texttt{doi:}\begingroup \urlstyle{tt}\Url}

\bibitem{uecms1}
\hrefCMSnoop {}{{CMS Collaboration}, ``Measurement of the underlying event
  activity at the {LHC} with {$\sqrt{s}= 7\TeV$} and comparison with
  {$\sqrt{s}= 0.9\TeV$}'',} \textit{ JHEP} \textbf{ 09} (2011) 109,
  \href{http://dx.doi.org/10.1007/JHEP09(2011)109}{\doi{10.1007/JHEP09(2011)109}},
\href{http://www.arXiv.org/abs/1107.0330}{\texttt{arXiv:1107.0330}}.

\bibitem{uecms2}
\hrefCMSnoop {}{{CMS Collaboration}, ``Measurement of the underlying event
  activity using charged-particle jets in proton-proton collisions at
  {$\sqrt{s}= 2.76\TeV$}'',} \textit{ JHEP} \textbf{ 09} (2015) 137,
  \href{http://dx.doi.org/10.1007/JHEP09(2015)137}{\doi{10.1007/JHEP09(2015)137}},
\href{http://www.arXiv.org/abs/1507.07229}{\texttt{arXiv:1507.07229}}.

\bibitem{uedycms}
\hrefCMSnoop {}{{CMS Collaboration}, ``Measurement of the underlying event in
  the {Drell--Yan} process in proton-proton collisions at {$\sqrt{s}=
  7\TeV$}'',} \textit{ Eur. Phys. J. C} \textbf{ 72} (2012) 2080,
  \href{http://dx.doi.org/10.1140/epjc/s10052-012-2080-4}{\doi{10.1140/epjc/s10052-012-2080-4}},
\href{http://www.arXiv.org/abs/1204.1411}{\texttt{arXiv:1204.1411}}.

\bibitem{uealice}
\hrefCMSnoop {}{{ALICE Collaboration}, ``Underlying event measurements in pp
  collisions at {$\sqrt{s}=0.9$} and {7\TeV} with the {ALICE} experiment at the
  {LHC}'',} \textit{ JHEP} \textbf{ 07} (2012) 116,
  \href{http://dx.doi.org/10.1007/JHEP07(2012)116}{\doi{10.1007/JHEP07(2012)116}},
\href{http://www.arXiv.org/abs/1112.2082}{\texttt{arXiv:1112.2082}}.

\bibitem{ueatlas1}
\hrefCMSnoop {}{{ATLAS Collaboration}, ``Measurement of underlying event
  characteristics using charged particles in pp collisions at {$\sqrt{s} =
  900\GeV$} and {7\TeV} with the {ATLAS} detector'',} \textit{ Phys. Rev. D}
  \textbf{ 83} (2011) 112001,
  \href{http://dx.doi.org/10.1103/PhysRevD.83.112001}{\doi{10.1103/PhysRevD.83.112001}},
\href{http://www.arXiv.org/abs/1012.0791}{\texttt{arXiv:1012.0791}}.

\bibitem{ueatlas2}
\hrefCMSnoop {}{{ATLAS Collaboration}, ``Measurements of underlying-event
  properties using neutral and charged particles in pp collisions at {900\GeV}
  and {7\TeV} with the {ATLAS} detector at the {LHC}'',} \textit{ Eur. Phys. J.
  C} \textbf{ 71} (2011) 1636,
  \href{http://dx.doi.org/10.1140/epjc/s10052-011-1636-z}{\doi{10.1140/epjc/s10052-011-1636-z}},
\href{http://www.arXiv.org/abs/1103.1816}{\texttt{arXiv:1103.1816}}.

\bibitem{ueatlas3}
\hrefCMSnoop {}{{ATLAS Collaboration}, ``Measurement of the dependence of
  transverse energy production at large pseudorapidity on the hard-scattering
  kinematics of proton-proton collisions at {$\sqrt{s} = 2.76\TeV$} with
  {ATLAS}'',} \textit{ Phys. Lett. B} \textbf{ 756} (2016) 10,
  \href{http://dx.doi.org/10.1016/j.physletb.2016.02.056}{\doi{10.1016/j.physletb.2016.02.056}},
\href{http://www.arXiv.org/abs/1512.00197}{\texttt{arXiv:1512.00197}}.

\bibitem{ueatlas4}
\hrefCMSnoop {}{{ATLAS Collaboration}, ``Measurement of distributions sensitive
  to the underlying event in inclusive {Z}-boson production in pp collisions at
  {$\sqrt{s}= 7\TeV$} with the {ATLAS} detector'',} \textit{ Eur. Phys. J. C}
  \textbf{ 74} (2014) 3195,
  \href{http://dx.doi.org/10.1140/epjc/s10052-014-3195-6}{\doi{10.1140/epjc/s10052-014-3195-6}},
\href{http://www.arXiv.org/abs/1409.3433}{\texttt{arXiv:1409.3433}}.

\bibitem{uecdf}
\hrefCMSnoop {}{{CDF} Collaboration, ``Studying the underlying event in
  {Drell--Yan} and high transverse momentum jet production at the
  {Tevatron}'',} \textit{ Phys. Rev. D} \textbf{ 82} (2010) 034001,
  \href{http://dx.doi.org/10.1103/PhysRevD.82.034001}{\doi{10.1103/PhysRevD.82.034001}},
\href{http://www.arXiv.org/abs/1003.3146}{\texttt{arXiv:1003.3146}}.

\bibitem{Sjostrand:1986ep}
\hrefCMSnoop {}{T.~Sj{\"o}strand and M.~Van~Zijl, ``A multiple interaction
  model for the event structure in hadron collisions'',} \textit{ Phys. Rev. D}
  \textbf{ 36} (1987) 2019,
  \href{http://dx.doi.org/10.1103/PhysRevD.36.2019}{\doi{10.1103/PhysRevD.36.2019}}.

\bibitem{Frankfurt:2011}
\hrefCMSnoop {}{L.~Frankfurt, M.~Strikman, and C.~Weiss, ``Transverse nucleon
  structure and diagnostics of hard parton-parton processes at {LHC}'',}
  \textit{ Phys. Rev. D} \textbf{ 83} (2011) 054012,
  \href{http://dx.doi.org/10.1103/PhysRevD.83.054012}{\doi{10.1103/PhysRevD.83.054012}},
  \href{http://www.arXiv.org/abs/1009.2559}{\texttt{arXiv:1009.2559}}.

\bibitem{Bansal:2016iri}
\hrefCMSnoop {}{R.~Kumar, M.~Bansal, S.~Bansal, and J.~B. Singh, ``{New
  observables for multiple-parton interactions measurements using Z+jets
  processes at the LHC}'',} \textit{ Phys. Rev. D} \textbf{ 93} (2016) 054019,
  \href{http://dx.doi.org/10.1103/PhysRevD.93.054019}{\doi{10.1103/PhysRevD.93.054019}},
\href{http://www.arXiv.org/abs/1602.05392}{\texttt{arXiv:1602.05392}}.

\bibitem{Alwall:2014hca}
J.~Alwall\hrefCMSnoop {}{ {et~al.}, ``{The automated computation of tree-level
  and next-to-leading order differential cross sections, and their matching to
  parton shower simulations}'',} \textit{ JHEP} \textbf{ 07} (2014) 079,
  \href{http://dx.doi.org/10.1007/JHEP07(2014)079}{\doi{10.1007/JHEP07(2014)079}},
  \href{http://www.arXiv.org/abs/1405.0301}{\texttt{arXiv:1405.0301}}.

\bibitem{Frixione:2007}
\hrefCMSnoop {}{S.~Frixione, P.~Nason, and C.~Oleari, ``Matching {NLO QCD}
  computations with parton shower simulations: the powheg method'',} \textit{
  JHEP} \textbf{ 11} (2007) 070,
  \href{http://dx.doi.org/10.1088/1126-6708/2007/11/070}{\doi{10.1088/1126-6708/2007/11/070}},
  \href{http://www.arXiv.org/abs/0709.2092}{\texttt{arXiv:0709.2092}}.

\bibitem{MINLO}
\hrefCMSnoop {}{K.~Hamilton, P.~Nason, and G.~Zanderighi, ``{MINLO}:
  multi-scale improved {NLO}'',} \textit{ JHEP} \textbf{ 10} (2012) 155,
  \href{http://dx.doi.org/10.1007/JHEP10(2012)155}{\doi{10.1007/JHEP10(2012)155}},
  \href{http://www.arXiv.org/abs/1206.3572}{\texttt{arXiv:1206.3572}}.

\bibitem{Maltoni:2003}
\hrefCMSnoop {}{F.~Maltoni and T.~Stelzer, ``{MadEvent}: automatic event
  generation with {MadGraph}'',} \textit{ JHEP} \textbf{ 02} (2003) 027,
  \href{http://dx.doi.org/10.1088/1126-6708/2003/02/027}{\doi{10.1088/1126-6708/2003/02/027}},
  \href{http://www.arXiv.org/abs/hep-ph/0208156}{\texttt{arXiv:hep-ph/0208156}}.

\bibitem{Alwall:2011}
\hrefCMSnoop {}{J.~Alwall {et~al.}, ``{MadGraph 5}: going beyond'',} \textit{
  JHEP} \textbf{ 06} (2011) 128,
  \href{http://dx.doi.org/10.1007/JHEP06(2011)128}{\doi{10.1007/JHEP06(2011)128}},
  \href{http://www.arXiv.org/abs/1106.0522}{\texttt{arXiv:1106.0522}}.

\bibitem{Sjostrand:2007gs}
\hrefCMSnoop {}{T.~Sj{\"o}strand, S.~Mrenna, and P.~Z. Skands, ``A brief
  introduction to {PYTHIA 8.1}'',} \textit{ Comput. Phys. Commun.} \textbf{
  178} (2008) 852,
  \href{http://dx.doi.org/10.1016/j.cpc.2008.01.036}{\doi{10.1016/j.cpc.2008.01.036}},
  \href{http://www.arXiv.org/abs/0710.3820}{\texttt{arXiv:0710.3820}}.

\bibitem{Khachatryan:2015pea}
\hrefCMSnoop {}{{CMS Collaboration}, ``{Event generator tunes obtained from
  underlying event and multiparton scattering measurements}'',} \textit{ Eur.
  Phys. J. C} \textbf{ 76} (2016) 155,
  \href{http://dx.doi.org/10.1140/epjc/s10052-016-3988-x}{\doi{10.1140/epjc/s10052-016-3988-x}},
\href{http://www.arXiv.org/abs/1512.00815}{\texttt{arXiv:1512.00815}}.

\bibitem{Ball:2014uwa}
\hrefCMSnoop {}{{NNPDF} Collaboration, ``{Parton distributions for the LHC Run
  II}'',} \textit{ JHEP} \textbf{ 04} (2015) 040,
  \href{http://dx.doi.org/10.1007/JHEP04(2015)040}{\doi{10.1007/JHEP04(2015)040}},
\href{http://www.arXiv.org/abs/1410.8849}{\texttt{arXiv:1410.8849}}.

\bibitem{Agostinelli:2002hh}
\hrefCMSnoop {}{{GEANT4} Collaboration, ``{GEANT4}---a simulation toolkit'',}
  \textit{ Nucl. Instrum. Meth. A} \textbf{ 506} (2003) 250,
\href{http://dx.doi.org/10.1016/S0168-9002(03)01368-8}{\doi{10.1016/S0168-9002(03)01368-8}}.

\bibitem{hpp}
M.~B{\"a}hr\hrefCMSnoop {}{ {et~al.}, ``Herwig++ physics and manual'',}
  \textit{ Eur. Phys. J. C} \textbf{ 58} (2008) 639,
  \href{http://dx.doi.org/10.1140/epjc/s10052-008-0798-9}{\doi{10.1140/epjc/s10052-008-0798-9}},
  \href{http://www.arXiv.org/abs/0803.0883}{\texttt{arXiv:0803.0883}}.

\bibitem{TRK-11-001}
\hrefCMSnoop {}{{CMS Collaboration}, ``{Description and performance of track
  and primary-vertex reconstruction with the CMS tracker}'',} \textit{ JINST}
  \textbf{ 9} (2014) P10009,
  \href{http://dx.doi.org/10.1088/1748-0221/9/10/P10009}{\doi{10.1088/1748-0221/9/10/P10009}},
\href{http://www.arXiv.org/abs/1405.6569}{\texttt{arXiv:1405.6569}}.

\bibitem{Chatrchyan:2012xi}
\hrefCMSnoop {}{{CMS Collaboration}, ``{Performance of CMS muon reconstruction
  in pp collision events at $\sqrt{s} = 7\TeV$}'',} \textit{ JINST} \textbf{ 7}
  (2012) P10002,
  \href{http://dx.doi.org/10.1088/1748-0221/7/10/P10002}{\doi{10.1088/1748-0221/7/10/P10002}},
\href{http://www.arXiv.org/abs/1206.4071}{\texttt{arXiv:1206.4071}}.

\bibitem{Chatrchyan:2008zzk}
\hrefCMSnoop {}{{CMS Collaboration}, ``The {CMS} experiment at the {CERN}
  {LHC}'',} \textit{ JINST} \textbf{ 3} (2008) S08004,
  \href{http://dx.doi.org/10.1088/1748-0221/3/08/S08004}{\doi{10.1088/1748-0221/3/08/S08004}}.

\bibitem{Sirunyan:2017ulk}
\hrefCMSnoop {}{{CMS Collaboration}, ``{Particle-flow reconstruction and global
  event description with the CMS detector}'',} \textit{ JINST} \textbf{ 12}
  (2017) P10003,
  \href{http://dx.doi.org/10.1088/1748-0221/12/10/P10003}{\doi{10.1088/1748-0221/12/10/P10003}},
\href{http://www.arXiv.org/abs/1706.04965}{\texttt{arXiv:1706.04965}}.

\bibitem{Bodek:2012id}
A.~Bodek\hrefCMSnoop {}{ {et~al.}, ``Extracting muon momentum scale corrections
  for hadron collider experiments'',} \textit{ Eur. Phys. J. C} \textbf{ 72}
  (2012) 2194,
  \href{http://dx.doi.org/10.1140/epjc/s10052-012-2194-8}{\doi{10.1140/epjc/s10052-012-2194-8}},
\href{http://www.arXiv.org/abs/1208.3710}{\texttt{arXiv:1208.3710}}.

\bibitem{CMS-PAPERS-TRG-12-001}
\hrefCMSnoop {}{{CMS Collaboration}, ``The {CMS} trigger system'',} \textit{ J.
  Instrum.} \textbf{ 12} (2017) P01020,
  \href{http://dx.doi.org/10.1088/1748-0221/12/01/P01020}{\doi{10.1088/1748-0221/12/01/P01020}}.

\bibitem{trk}
\hrefCMSnoop {}{{CMS Collaboration}, ``{CMS} tracking performance results from
  early {LHC} operation'',} \textit{ Eur. Phys. J. C} \textbf{ 70} (2010) 1165,
  \href{http://dx.doi.org/10.1140/epjc/s10052-010-1491-3}{\doi{10.1140/epjc/s10052-010-1491-3}},
\href{http://www.arXiv.org/abs/1007.1988}{\texttt{arXiv:1007.1988}}.

\bibitem{unfold}
\hrefCMSnoop {}{G.~D'Agostini, ``{A multidimensional unfolding method based on
  Bayes' theorem}'',} \textit{ Nucl. Instrum. Meth. A} \textbf{ 362} (1995)
  487,
  \href{http://dx.doi.org/10.1016/0168-9002(95)00274-X}{\doi{10.1016/0168-9002(95)00274-X}}.

\bibitem{CMS:2010mua}
\href {http://inspirehep.net/record/925300}{{CMS Collaboration}, ``Measurement
  of tracking efficiency'',} CMS Physics Analysis Summary CMS-PAS-TRK-10-002,
  2010.

\end{thebibliography}\endgroup

\cleardoublepage \appendix\section{The CMS Collaboration \label{app:collab}}\begin{sloppypar}\hyphenpenalty=5000\widowpenalty=500\clubpenalty=5000\vskip\cmsinstskip
\textbf{Yerevan~Physics~Institute,~Yerevan,~Armenia}\\*[0pt]
A.M.~Sirunyan, A.~Tumasyan
\vskip\cmsinstskip
\textbf{Institut~f\"{u}r~Hochenergiephysik,~Wien,~Austria}\\*[0pt]
W.~Adam, F.~Ambrogi, E.~Asilar, T.~Bergauer, J.~Brandstetter, E.~Brondolin, M.~Dragicevic, J.~Er\"{o}, M.~Flechl, M.~Friedl, R.~Fr\"{u}hwirth\cmsAuthorMark{1}, V.M.~Ghete, J.~Grossmann, J.~Hrubec, M.~Jeitler\cmsAuthorMark{1}, A.~K\"{o}nig, N.~Krammer, I.~Kr\"{a}tschmer, D.~Liko, T.~Madlener, I.~Mikulec, E.~Pree, D.~Rabady, N.~Rad, H.~Rohringer, J.~Schieck\cmsAuthorMark{1}, R.~Sch\"{o}fbeck, M.~Spanring, D.~Spitzbart, J.~Strauss, W.~Waltenberger, J.~Wittmann, C.-E.~Wulz\cmsAuthorMark{1}, M.~Zarucki
% ...
\vskip\cmsinstskip
% ...
1:~Also at~Vienna~University~of~Technology,~Vienna,~Austria\\
% ...
\end{sloppypar}
\end{document}